\documentclass[conference,compsoc]{IEEEtran}

\ifCLASSOPTIONcompsoc
  \usepackage[nocompress]{cite}
\else
  \usepackage{cite}
\fi

\IEEEoverridecommandlockouts

\usepackage{acronym}
\usepackage{setspace}
\usepackage{longtable}
\usepackage[figuresright]{rotating}
\usepackage{epsfig,endnotes,color}
\usepackage{graphicx}
\usepackage{amsmath}
\usepackage{caption}
\usepackage{subcaption}
\usepackage{url}

\usepackage{CJKutf8}
\usepackage{lscape, latexsym, amssymb, algorithmic, multirow}
\usepackage[linesnumbered, vlined, ruled]{algorithm2e}
\usepackage{multirow}
\usepackage{mathtools, bbm, color}
\usepackage{booktabs}
\usepackage{amsthm,mathrsfs,amsfonts,dsfont}
\usepackage{hhline}
\usepackage{tabu}
\usepackage{colortbl}
\usepackage{enumerate}

\usepackage[graphicx]{realboxes}
\usepackage{float}
\usepackage{tikz}
\newcommand*\emptcirc[1][1ex]{\tikz\draw (0,0) circle (#1);} 

\newcommand*\fullcirc[1][1ex]{\tikz\fill (0,0) circle (#1);} 

\usepackage{footnote}
\usepackage{color}
\usepackage{colortbl}
\usepackage{pifont}
\usepackage{bbding}
\usepackage{threeparttable}
\usepackage{rotating}
\usepackage{xcolor}
\usepackage{tcolorbox}
\definecolor{shadecolor}{rgb}{0.92,0.92,0.92}
\newtcolorbox{mybox}{ colframe=black,colback=gray!15,boxrule=1pt,arc=2pt,left=2pt,right=2pt,top=1pt,bottom=1pt}

\usepackage[misc]{ifsym}
\usepackage{listings}
\definecolor{codegreen}{rgb}{0,0.6,0}
\definecolor{codegray}{rgb}{0.5,0.5,0.5}
\definecolor{codepurple}{rgb}{0.58,0,0.82}
\definecolor{backcolour}{rgb}{0.95,0.95,0.95}
\lstdefinestyle{mystyle}{
 commentstyle=\color{codegreen},
 keywordstyle=\color{magenta},
 numberstyle=\tiny\color{codegray},
 stringstyle=\color{codepurple},
 basicstyle=\footnotesize\scriptsize,
 breakatwhitespace=false,         
 breaklines=true,                 
 captionpos=b,                    
 keepspaces=true,
 numbers=left,  
 showspaces=false,                
 showstringspaces=false,
 showtabs=false,                  
 tabsize=1,
 xleftmargin=\parindent,
}
\def\BibTeX{{\rm B\kern-.05em{\sc i\kern-.025em b}\kern-.08em
    T\kern-.1667em\lower.7ex\hbox{E}\kern-.125emX}}

\def\BibTeX{{\rm B\kern-.05em{\sc i\kern-.025em b}\kern-.08em
    T\kern-.1667em\lower.7ex\hbox{E}\kern-.125emX}}

\newcommand{\system}{\textsc{SyzTrust}\xspace}
\newcommand{\sysbasic}{\textsc{SyzTrust\_Basic}\xspace}
\newcommand{\sysstate}{\textsc{SyzTrust\_State}}
\newcommand{\sysfstate}{\textsc{SyzTrust\_FState}\xspace}

\newcommand{\syzkaller}{\textsc{Syzkaller}\xspace}

\newif\ifcommentcond
\commentcondfalse 

\newif\ifupdatecond
\updatecondfalse

\newif\ifparasumcond
\parasumcondfalse

\newcounter{thechalls}
\makeatletter
\@ifpackageloaded{cleveref}{\crefname{thechalls}{Challenge}{Challenges}}{}
\makeatother

\newcommand{\ignore}[1]{}

\newcommand{\bugCount}{70} 
\newcommand{\bugconfirm}{28}
\newcommand{\higherCodeCov}{66} 
\newcommand{\higherStateCov}{651}
\newcommand{\higherBug}{31}

\newcommand{\cve}{10}
\newcommand{\etmtracingspeed}{168 KB}

\newcommand{\syzkallercode}{1,191}
\newcommand{\syzbasiccode}{1,983}

\newcommand{\syzkallerbug}{16} 
\newcommand{\fstatebug}{21}
\newcommand{\syzbasicbug}{20}

\newcommand{\stateinfeacc}{83.3}

\newcommand{\linkteebug}{19}
\newcommand{\linkteecov}{10,710}
\newcommand{\linkteestate}{182,324}
\newcommand{\mtowerbug}{38}
\newcommand{\mtowercov}{2,105}
\newcommand{\mtowerstate}{3,994}
\newcommand{\tinyteebug}{13}
\newcommand{\tinyteecov}{1,072}
\newcommand{\tinyteestate}{2,908}

\acrodef{AFL}{American Fuzzy Lop}
\acrodef{APSR}{Application Program Status Register}
\acrodef{ASLR}{Address Space Layout Randomization}
\acrodef{CFG}{Control Flow Graph}
\acrodef{COTS}{Commercial-off-the-shelf}
\acrodef{CPU}{Central Processing Unit}
\acrodef{CPSR}{Current Program Status Register}
\acrodef{CWE}{Common Weakness Enumeration}
\acrodef{DEP}{Data Execution Prevention}
\acrodef{DMA}{Direct Memory Access}
\acrodef{DoS}{Denial-of-Service}
\acrodef{DSE}{Dynamic Symbolic Execution}
\acrodef{HAL}{Hardware Abstraction Layer}
\acrodef{HLM}{High-level Modeling}
\acrodef{ICS}{Industrial Control System}
\acrodef{I/O}{Input/Output}
\acrodef{IoT}{Internet of Things}
\acrodef{IR}{Intermediate Representation}
\acrodef{ISA}{Instruction Set Architecture}
\acrodef{ISR}{Interrupt Service Routine}
\acrodef{LSB}{Least Significant Bit}
\acrodef{MMIO}{Memory-mapped Input/Output}
\acrodef{MCU}{Microcontroller Unit}
\acrodef{MMU}{Memory Management Unit}
\acrodef{MPU}{Memory Protection Unit}
\acrodef{NVIC}{Nested Vectored Interrupt Controller}
\acrodef{OS}{Operating System}
\acrodef{OSS}{Open-Source Software}
\acrodef{PC}{Personal Computer}
\acrodef{PLC}{Programmable Logic Controller}
\acrodef{PUT}{Program Under Test}
\acrodef{RAID}{Redundant Array of Independent Disks}
\acrodef{RISC}{Reduced Instruction Set Computing}
\acrodef{RTOS}{Real-Time Operating System}
\acrodef{SBI}{Static Binary Instrumentation}
\acrodef{SDK}{Software Development Kit}
\acrodef{SMT}{Simultaneous Multithreading}
\acrodef{SoC}{System-on-Chip}
\acrodef{TDP}{Thermal Design Power}
\acrodef{UART}{Universal asynchronous receiver-transmitter}
\acrodef{OP-TEE}{Open Portable Trusted Execution Environment}
\acrodef{OEM}{Original Equipment Manufacturer}
\acrodef{TA}{Trusted Application}
\acrodef{CA}{Client Application}
\acrodef{Trusted OS}{Trusted Operating System}
\acrodef{TEE}{Trusted Execution Environment}
\acrodef{DWT}{Data Watchpoint and Trace Unit}
\acrodef{LCSAJ}{Liner Code Sequence and Jump}

\begin{document}

\title{SyzTrust: State-aware Fuzzing on Trusted OS Designed for IoT Devices}

\author{
\IEEEauthorblockN {Qinying Wang\IEEEauthorrefmark{1}\IEEEauthorrefmark{2}, 
Boyu Chang\IEEEauthorrefmark{1},
Shouling Ji\IEEEauthorrefmark{1}$^{(\textrm{\Letter}),}$\thanks{Shouling Ji is the corresponding author.},
Yuan Tian\IEEEauthorrefmark{3},
Xuhong Zhang\IEEEauthorrefmark{1},
Binbin Zhao\IEEEauthorrefmark{4},
Gaoning Pan\IEEEauthorrefmark{1}, \\ 
Chenyang Lyu\IEEEauthorrefmark{1},
Mathias Payer\IEEEauthorrefmark{2}, 
Wenhai Wang\IEEEauthorrefmark{1}, 
Raheem Beyah\IEEEauthorrefmark{4}
}

\IEEEauthorblockA {
  \IEEEauthorrefmark{1}Zhejiang University,
  \IEEEauthorrefmark{2}EPFL,
  \IEEEauthorrefmark{3}University of California, Los Angelos,
  \IEEEauthorrefmark{4}Georgia Institute of Technology \\
}

\IEEEauthorblockA {
  E-mails: \{wangqinying, bychang, sji\}@zju.edu.cn, yuant@ucla.edu, zhangxuhong@zju.edu.cn, binbin.zhao@gatech.edu, \\
  \{pgn, puppet\}@zju.edu.cn, mathias.payer@nebelwelt.net, zdzzlab@zju.edu.cn, rbeyah@ece.gatech.edu
}
}

\maketitle

\thispagestyle{plain}
\pagestyle{plain}
\bstctlcite{IEEEexample:BSTcontrol}

\begin{abstract}

Trusted Execution Environments (TEEs) embedded in IoT devices provide a deployable solution to secure IoT applications at the hardware level.
By design, in TEEs, the Trusted Operating System (Trusted OS) is the primary component. It enables the TEE to use security-based design techniques, such as data encryption and identity authentication.
Once a Trusted OS has been exploited, the TEE can no longer ensure security.
However, Trusted OSes for IoT devices have received little security analysis, which is challenging from several perspectives: 
(1) Trusted OSes are closed-source and have an unfavorable environment for sending test cases and collecting feedback. 
(2) Trusted OSes have complex data structures and require a stateful workflow, which limits existing vulnerability detection tools.

To address the challenges, we present \system{}, the first state-aware fuzzing framework for vetting the security of resource-limited Trusted OSes. 
\system{} adopts a hardware-assisted framework to enable fuzzing Trusted OSes directly on IoT devices as well as tracking state and code coverage non-invasively.
\system{} utilizes composite feedback to guide the fuzzer to effectively explore more states as well as to increase the code coverage.
We evaluate \system{} on Trusted OSes from three major vendors: Samsung, Tsinglink Cloud, and Ali Cloud.
These systems run on Cortex M23/33 MCUs, which provide the necessary abstraction for embedded TEEs.
We discovered \bugCount{} previously unknown vulnerabilities in their Trusted OSes, receiving \cve{} new CVEs so far.
Furthermore, compared to the baseline, \system{} has demonstrated significant improvements, including \higherCodeCov{}\% higher code coverage, \higherStateCov{}\% higher state coverage, and \higherBug{}\% improved vulnerability-finding capability.
We report all discovered new vulnerabilities to vendors and open source \system{}.  
\end{abstract}

\IEEEpeerreviewmaketitle

%%%%%%%%%%%%%%%%%%%%%%%%%%%%%%%%%%%%%%%%%%%%%%%%%%%%%%%%%%%%%%%%%%%%%%%%%%%%%%%%
\section{Introduction}
%%%%%%%%%%%%%%%%%%%%%%%%%%%%%%%%%%%%%%%%%%%%%%%%%%%%%%%%%%%%%%%%%%%%%%%%%%%%%%%%
Trusted Execution Environments (TEEs) are essential to securing important data and operations in IoT devices. 
GlobalPlatform, the leading technical standards organization, has reported a 25-percent increase in the number of TEE-enabled IoT processors being shipped quarterly, year-over-year \cite{TEEBillion}.
Recently, major IoT vendors such as Samsung have designed TEEs for low-end Microcontroller Units (MCUs) \cite{mTower, TinyTEE, LinkTEE} and device manufacturers have embedded the TEE in IoT devices such as unmanned aerial vehicles and smart locks, to protect sensitive data and to provide key management services.
A TEE is composed of Client Applications (CAs), Trusted Applications (TAs), and a Trusted Operating System (Trusted OS).
Among them, the Trusted OS is the primary component to enable the TEE using security-based design techniques, and its security is the underlying premise of a reliable TEE where the code and data are protected in terms of confidentiality and integrity.
Unfortunately, implementation flaws in Trusted OSes violate the protection guarantees, bypassing confidentiality and integrity guarantees.
These flaws lead to critical consequences, including sensitive information leakage (CVE-2019-25052) and code execution within the Trusted OS context\cite{luo2020runtime, TEEEscalation}. 
Once attackers gain control of Trusted OSes, they can launch critical attacks,
such as creating a backdoor to the Linux kernel \cite{LinuxKernelBackdoor}, and extracting full disk encryption keys of Android's KeyMaster service \cite{breakkeymaster}.
  
While TEEs are increasingly embedded in IoT devices, the security of Trusted OS for IoT devices remains under studied. 
Considering the emerging amount of diversified MCUs and IoT devices, manual analysis, such as reverse engineering, requires significant expert efforts and is therefore infeasible at scale. 
Recent academic works use fuzzing to automate TEE testing.
However, unlike Trusted OSes for Android devices, Trusted OSes for IoT devices are built on TrustZone-M with low-power and cost-sensitive MCUs, including NuMicro M23.
Thus, Trusted OSes for IoT devices are more hardware-dependent and resource-constrained, complicating the development of scalable and usable testing approaches with different challenges.
In the following, we conclude two challenges for fuzzing IoT Trusted OSes.

\noindent\textbf{Challenge \uppercase\expandafter{\romannumeral1}: The inability of instrumentation and restricted environment.}
\label{cha:1} 
Most Trusted OSes are closed-source.  
Additionally, TEE implementations, especially the Trusted OSes are often encrypted by IoT vendors, which implies the inability to instrument and monitor the code execution in the secure world.  
Accordingly, classic feedback-driven fuzzing cannot be directly applied to the scenario of testing TEEs including TAs and Trusted OSes.
Existing works either rely on on-device binary instrumentations \cite{buschteezz} or require professional domain knowledge and rehosting through proprietary firmware emulation \cite{harrison2020partemu} to enable testing and coverage tracking.
However, as for the Trusted OSes designed for IoT devices, the situation is more challenging due to the following two reasons.
First, IoT devices are severely resource-limited, while existing binary instrumentations are heavy-weight for them and considerably limit their execution speed.
Second, as for rehosting, IoT devices are mostly hardware-dependent, rendering the reverse engineering and implementation effort for emulated software and hardware components unacceptable. 
In addition, rehosting faces the limitation of the inaccuracy of modeling the behavior of hardware components. 
To our best knowledge, the only existing TEE rehosting solution PartEmu \cite{harrison2020partemu} is not publicly available and does not support the mainstream TEE based on Cortex-M MCUs designed for IoT devices.

\noindent\textbf{Challenge \uppercase\expandafter{\romannumeral2}: Complex structure and stateful workflow.}
Trusted OSes for IoT devices are surprisingly complex.
Specifically, Trusted OSes implement multiple cryptographic algorithms, such as AES and MAC, without underlying hardware support for these algorithms as would be present on Cortex-A processors.
To implement these algorithms in a secure way, Trusted OSes maintain several state diagrams to store the execution contexts and guide the execution workflow.
To explore more states of a Trusted OS, a fuzzer needs to feed syscall sequences in several specific orders with different specific state-related argument values.
Without considering such statefulness of Trusted OSes, coverage-based fuzzers are unlikely to explore further states, causing the executions to miss the vulnerabilities hidden in a deep state. 
Unfortunately, existing fuzzing techniques lack state-awareness for Trusted OSes. 
Specifically, they have trouble understanding which state a Trusted OS reaches since there are no rich-semantics response codes to indicate it. 
In addition, due to the lack of source code and the inability of instrumentation, it is hard to infer and extract the state variables by program analysis.

\noindent\textbf{Our solution.}
To address the above key challenges, we propose and implement \system{}, the first fuzzing framework targeting Trusted OSes for IoT devices, supporting state and coverage-guided fuzzing.
Specifically, we propose an on-device fuzzing framework and leverage a hardware-in-the-loop approach.
To support in-depth vulnerability detection, we propose a composite feedback mechanism that guides the fuzzer to explore more states and increase code coverage.

\system{} necessitates diverse contributions. 
First, to tackle Challenge \uppercase\expandafter{\romannumeral1}, we propose a hardware-assisted fuzzing framework to execute test cases as well as collect code coverage feedback.
Specifically, we decouple the execution engine from the rest of the fuzzer to enable directly executing test cases in the protective TEE secure world on the resource-limited MCU. 
To support coverage tracking, we present a selective trace collection approach instead of costly code instrumentation to enable tracing instructions on a target MCU.
In particular, we leverage the ARM Embedded Trace Macrocell (ETM) feature to collect raw trace packets by monitoring instruction and data buses on MCU with a low-performance impact.
However, the Trusted OS for IoT devices is resource constrained, which makes
storing ETM traces on board difficult and limits the fuzzing speed. 
Additionally, the TEE internals are complicated and have multiple components, which generate noisy trace packets.
Therefore, we offload heavy-weight tasks to a PC and carefully scheduled the fuzzing subprocesses in a more parallel way.
We also present an event- and address-based trace filter to handle the noisy trace packets that are not executed by the Trusted OS.
Additionally, we adopt an efficient code coverage calculation algorithm directly on the raw packets.

Second, as for the Challenge \uppercase\expandafter{\romannumeral2}, the vulnerability detection capability of coverage-based fuzzers is limited, and a more effective fuzzing strategy is required. 
Therefore, we propose a composite feedback mechanism, which enhances code coverage with state feedback.
\iffalse
As for collecting the state feedback, it is challenging to identify the internal states of a Trusted OS.
In our work, we utilize state variables that control the execution contexts to present the states of a Trusted OS.
\fi
To be specific, we utilize state variables that control the execution contexts to present the states of a Trusted OS.
However, such state variables are usually stored in closed-source and customized data structures within Trusted OSes.
Existing state variable inference methods either use explicit protocol packet sequences \cite{mcmahon2022closer} or require source codes of target software \cite{kim2022fuzzusb,281444}, which are unavailable for Trusted OSes.
Therefore, to identify the state-related members from those complex data structures, \system{} collects some heuristics for Trusted OS and utilizes them to perform an active state variable inference algorithm.
After that, \system{} monitors the state variable values during the fuzzing procedure as the state feedback.   

Finally, \system{} is the first end-to-end solution capable of fuzzing Trusted OSes for IoT devices.
Moreover, the design of the on-device fuzzing framework and modular implementation make \system{} more extensible. 
With several MCU-specific configurations, \system{} scales to Trusted OSes on different MCUs from different vendors.

\noindent\textbf{Evaluation.} 
We evaluate \system{} on real-world Trusted OSes from three leading IoT vendors Samsung, Tsinglink Cloud, and Ali Cloud. 
The evaluation result shows that \system{} is effective at discovering new vulnerabilities and exploring new states and codes.
As a result, \system{} has discovered \bugCount{} new vulnerabilities. 
Among them, vendors confirmed \bugconfirm{}, and assigned \cve{} CVE IDs. The vendors are still investigating others.
Compared to state-of-the-art approaches, \system{} finds more vulnerabilities, hits \higherCodeCov{}\% higher code branches, and \higherStateCov{}\% higher state coverage.

\noindent \textbf{Summary and contributions.} 

$\bullet$ 
We propose \system{}, the first fuzzing framework targeting Trusted OSes for IoT devices, supporting effective state and code coverage guided fuzzing. 
    With a carefully designed hardware-assisted fuzzing framework and a composite feedback mechanism, \system{} is extensible and configurable to different IoT devices.

$\bullet$ 
With \system{}, we evaluate three popular Trusted OSes on three leading IoT vendors and detect several previously unknown bugs.
    We have responsibly reported these vulnerabilities to the vendors and got acknowledged from vendors such as Samsung.
    We release \system{} as an open-source tool for facilitating further studies at \url{https://github.com/SyzTrust}.

%%%%%%%%%%%%%%%%%%%%%%%%%%%%%%%%%%%%%%%%%%%%%%%%%%%%%%%%%%%%%%%%%%%%%%%%%%%%%%%%
\section{Background}
%%%%%%%%%%%%%%%%%%%%%%%%%%%%%%%%%%%%%%%%%%%%%%%%%%%%%%%%%%%%%%%%%%%%%%%%%%%%%%%
\subsection{TEE and Trusted OS}
A TEE is a secure enclave on a device's main processor that is separated from the main OS. 
It ensures the confidentiality and integrity of code and data loaded inside it \cite{TEEWiki}.
For standardizing the TEE implementations, GlobalPlatform (GP) has developed a number of specifications. 
For instance, it specifies the TEE Internal Core API implemented in the Trusted OS to enable a TA to perform its security functions \cite{TEECoreAPI}.
However, it is difficult for vendors to implement a Trusted OS correctly since there are lots of APIs with complex and stateful functions defined in the GP TEE specifications.
For instance, the TEE Internal Core API defines six types of APIs, including cryptographic operations APIs supporting more than 20 complex cryptographic algorithms.
In addition, the TEE Internal Core API also requires that a Trusted OS shall implement state diagrams to manage the operation.
The two aforementioned factors make it challenging for device vendors to develop a secure Trusted OS.

\begin{figure}[tbp]
    \centering
    \includegraphics[width=0.4\textwidth]{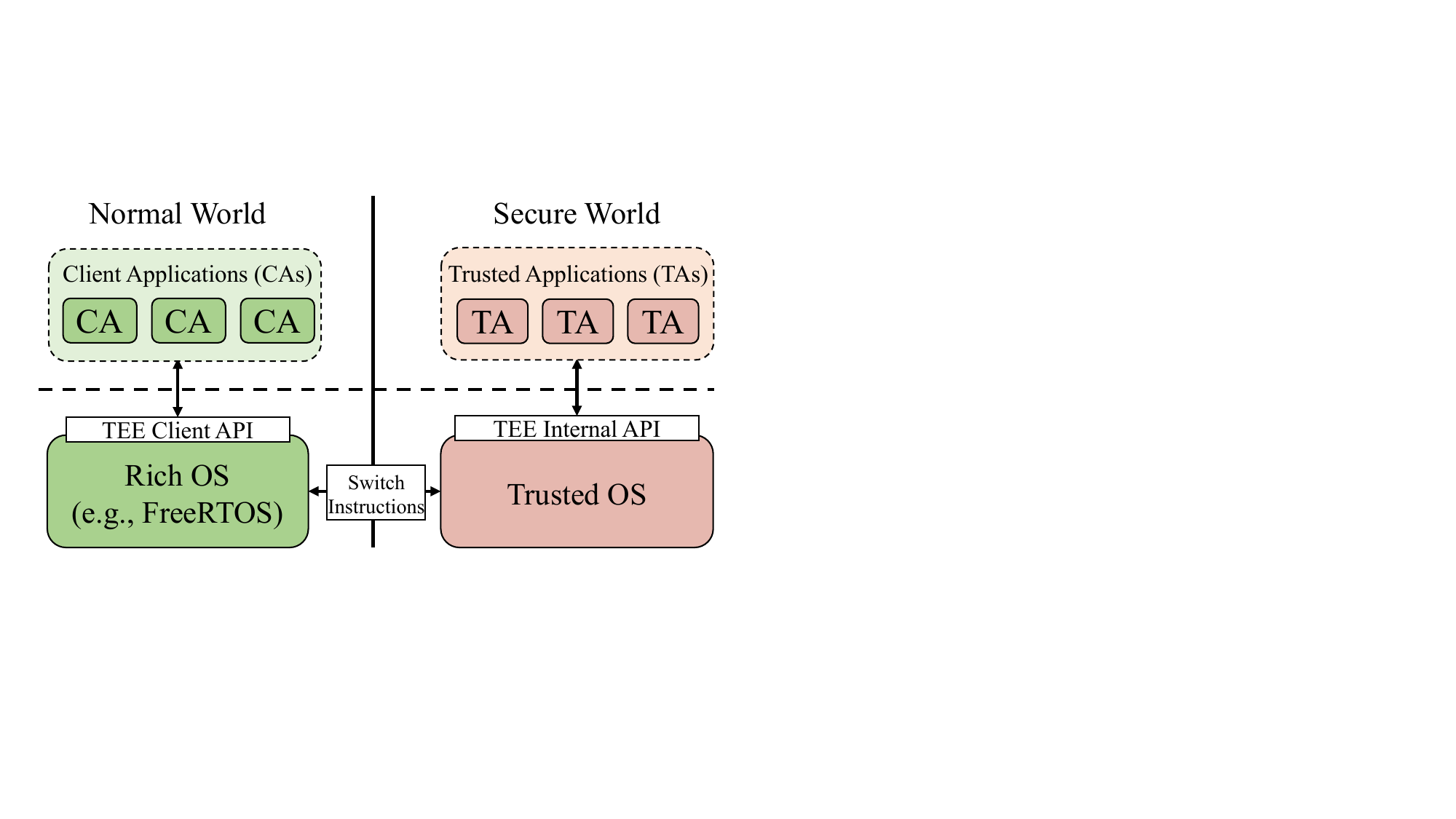} 
    \caption{Structure of TrustZone-M based TEE.}
    \label{fig:teearchitecture} 
  \end{figure}

ARM TrustZone has become the de-facto hardware technology to implement TEEs in mobile environments \cite{cerdeira2020sok} and has been deployed in servers \cite{hua2017vtz}, low-end IoT devices \cite{asokan2018assured, pinto2017iioteed}, and industrial control systems\cite{fitzek2015andix}.
For IoT devices, the Cortex-M23/33 MCUs, introduced by the ARM Community in 2016, are built on new TrustZone for ARMv8-M as security foundations for billions of devices \cite{CortexM23}.
TrustZone for ARMv8-M Cortex-M has been optimized for faster context switch and low-power applications and is designed from the ground up instead of being reused from Cortex-A \cite{pinto2019demystifying}.
As Figure \ref{fig:teearchitecture} shows, instead of utilizing a secure monitor in TrustZone for Cortex-A, the division of the secure and normal world is memory-based, and transitions take place automatically in the exception handle mode.
Based on the TrustZone-M, IoT vendors provide Trusted OS binaries to the device manufacturers and then the device manufacturers produce devices with device-specific TAs for the end users.
This paper focuses on the Trusted OSes from different IoT vendors and provides security insights for device manufacturers and end users.

\subsection{Debug Probe}
A debug probe is a special hardware device for low-level control of ARM-based MCUs, using DAP (Debug Access Port) provided by the ARM CoreSight Architecture \cite{armcoresight}.
It bridges the connection between a computer and an MCU and provides full debugging functionality, including watchpoints, flash memory breakpoints, memory, as well as register examination or editing.
In addition, a debug probe can record data and instruction accesses at runtime through the ARM ETM feature.
ETM is a subsystem of ARM Coresight Architecture and allows for traceability, whose function is similar to Intel PT.
The ETM generates \texttt{trace elements} for executed signpost instructions that enable reconstruction of all the executed instructions.
Utilizing the above features, the debug probe has shown its effectiveness in tracing and debugging malware \cite{ning2017ninja}, unpacking Android apps \cite{xue2021happer}, or fuzzing Linux peripheral drivers \cite{li2022mu}.

%%%%%%%%%%%%%%%%%%%%%%%%%%%%%%%%%%%%%%%%%%%%%%%%%%%%%%%%%%%%%%%%%%%%%%%%%%%%%%%%
  \section{Threat Model}
  \label{section: threatmodel}
  %%%%%%%%%%%%%%%%%%%%%%%%%%%%%%%%%%%%%%%%%%%%%%%%%%%%%%%%%%%%%%%%%%%%%%%%%%%%%%%%
  Our attacker tries to achieve multiple goals: gaining control over, extracting confidential information from, and causing crashes in other Trusted Applications (TAs) hosted on the same Trusted OS or the Trusted OS itself. 
  We consider two practical attack scenarios. First, an attacker can exploit our discovered vulnerabilities by providing carefully crafted data to a TA. They can utilize a malicious Client Application (CA) to pass the crafted data to a TA. 
  For instance, in mTower, CVE-2022-38511 (ID 1 in Table~\ref{tab:allbugs}) can be triggered by passing a large key size value from a CA to a TA. 
  Second, an attacker can exploit our discovered vulnerabilities by injecting a malicious TA into the secure world. 
  They can do this through rollback attacks or electromagnetic fault injections (CVE-2022-47549).

\begin{figure*}[tbp]
  \centering
  \includegraphics[width=0.85\textwidth]{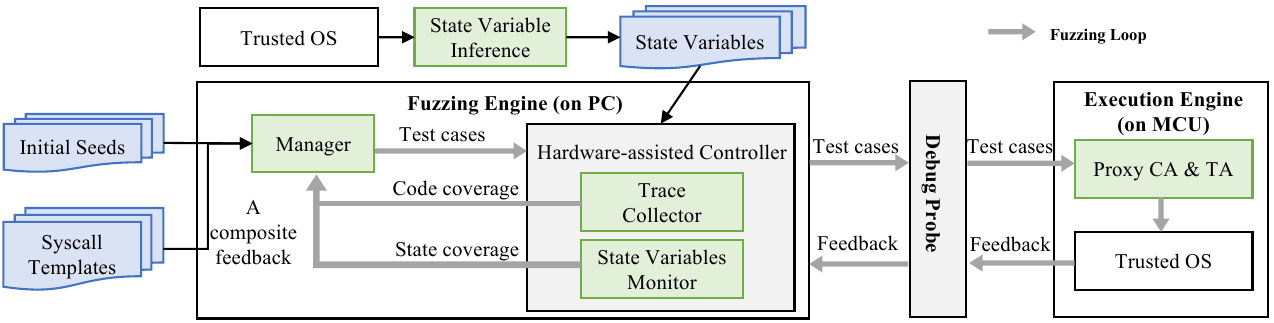} 
  \caption{Overview of \system{}. \system{} consists of a fuzzing engine running on a PC, an execution engine running on an MCU, and a debug probe to bridge the fuzzing engine and the execution engine.
  }
  
  \vspace{-0.2in}
  \label{fig:overview} 
\end{figure*}
%%%%%%%%%%%%%%%%%%%%%%%%%%%%%%%%%%%%%%%%%%%%%%%%%%%%%%%%%%%%%%%%%%%%%%%%%%%%%%%%
\section{Design}
\label{section: DesignImplementation}
%%%%%%%%%%%%%%%%%%%%%%%%%%%%%%%%%%%%%%%%%%%%%%%%%%%%%%%%%%%%%%%%%%%%%%%%%%%%%%%%

Figure~\ref{fig:overview} gives an overview of \system{}'s design. \system{} includes two modules: the \emph{fuzzing engine} on the Personal Computer (PC) and the \emph{execution engine} on the MCU. 
The fuzzing engine generates and sends test cases to the MCU via the debug probe.
The execution engine executes the received test case on the target Trusted OS. 

At a high level, we propose a hardware-assisted fuzzing framework and a composite feedback mechanism to guide the fuzzer.
Given the inaccessible environment of Trusted OSes, we design a TA and CA pair as a proxy to the Trusted OS and utilize a debug probe to access the MCU for feedback collection.
To handle the challenge of limited resources, we decouple the execution engine from \system{} and only run it on the MCU.
This allows \system{}, with its resource-demanding core components, to run more effectively on a PC.
To handle the statefulness of Trusted OSes, we include state feedback with code coverage in the composite feedback. State variables represent internal states, and our inference method identifies them in closed-source Trusted OSes.

The main workflow is as follows. 
First, \system{} accepts two inputs, including initial seeds and syscall templates.
Second, the manager generates test cases through two fuzzing tasks, including generating a new test case from scratch based on syscall templates or by mutating a selected seed.
Then, the generated test cases are delivered to the execution engine on MCU through the debug probe.
The execution engine executes these test cases to test the Trusted OS.
Meanwhile, the debug probe selectively records the executed instruction trace, which is processed by our trace collector as an alternative to code coverage.
Additionally, the state variable monitor tracks the values of a set of target state variables to calculate state coverage via the debug probe.
Finally, the code and state coverage is fed to the manager as composite feedback to guide the fuzzer.
The above procedures are iteratively executed and we denote the iterative workflow as the fuzzing loop.

\subsection{Inputs}\label{sec:input}
Syscall templates and initial seeds are fed to \system{}.

\textbf{Syscall templates.} Provided syscall templates, \system{} is more likely to generate valid test cases by following the format defined by the templates. 
Syscall templates define the syscalls with their argument types as well as special types that provide the semantics of an argument. 
For instance, \texttt{resources} represent the values that need to be passed from an output of one syscall to an input of another syscall, which refers to the data dependency between syscalls. 
We manually write syscall templates for the Internal Core API for the GP TEE in Syzlang (\syzkaller's syscall description language~\cite{syzkaller}).
However, a value with a complex structure type, which is common in Trusted OSes, cannot be used as a resource value due to the design limitation of Syzlang.
To handle this limitation, we extend the syscall templates with helper syscalls, which accept values of complex structures and output their point addresses as \texttt{resource} values.
In total, we have 37 syscall templates and 8 helper syscalls, covering all the standard cryptographic APIs.

\textbf{Initial seeds.} Since syscall templates are used to generate syntactically valid inputs, \system{} requires initial seeds to collect some valid syscall sequences and arguments and speed up the fuzzing  procedure. Provided initial seeds, \system{} continuously generates test cases to test a target Trusted OS by mutating them.
However, there is no off-the-shelf seed corpus for testing Trusted OSes, and constructing one is not trivial, as the syscall sequences with arguments should follow critical crypto constraints, and manually constructing valid sequences will need much effort.
For example, a valid seed includes a certain order of syscall sequences to initiate an encryption operation, and the key and the encryption mode should be consistent with supporting encryption.
Fortunately, OP-TEE \cite{optee}, an open-source TEE implementation for Cortex-A, offers a test suite, which can be utilized to construct seeds for most Trusted OSes.
Specifically, we automatically inject codes in TAs provided by the test suite to log the syscall names and their arguments. 
Then we automatically convert those logs into seeds following the format required by \system{}.
In addition, we automatically add data dependencies between syscalls in the seed corpus. 
Accordingly, we identify the return values from syscalls that are input arguments of other syscalls and add helper syscalls for the identified values.

Although the syscall template and seed corpus construction require extra work, it is a one-shot effort and can be used in the further study of Trusted OS security.

\subsection{A Hardware-assisted Framework}\label{sec:fuzzingframework}
Now our hardware-assisted fuzzing framework enables the generated test cases to be executed in protective and resource-constrained trusted environments on IoT devices.
To handle the constrained resources in Challenge \uppercase\expandafter{\romannumeral1}, we decouple the execution engine from the other components of \system to ease the execution overhead of the MCU.
As shown in Figure \ref{fig:overview}, the fuzzing engine, which includes most core components of \system and requires more computing resources and memory space, runs on a PC, while only the execution engine runs on the MCU.
Thus, \system does the heavy tasks, including seed preservation, seed selection, type-aware generation, and mutation.

As for the protected execution environment in Challenge \uppercase\expandafter{\romannumeral1}, we design a pair of CA and TA as the execution engine that executes the test cases to test the Trusted OSes.
We utilize a debug probe to bridge the connection between the fuzzing engine and the execution engine.
In the following, we introduce the design of delivering test cases and the pair of CA and TA.
First, as shown in Figure \ref{fig:overview}, the debug probe transfers test cases and feedback between the fuzzing engine and the execution engine.
Specifically, a debug probe can directly access the memory of the target MCU.
Thus, the debug probe accepts generated test cases from the manager engine and writes them to a specific memory of the target MCU. 
As for test case transportation, we design a serializer to encode the test cases into minimum binary data denoted as payloads before sending them to the MCU.
A payload includes a sequence of syscalls with their syscall names, syscall argument types, and syscall argument values.
In addition, \system{} denotes syscall names as ordinal numbers to minimize the transportation overhead.
Second, the pair of CA and TA plays the role of a proxy to handle the test cases.
Accordingly, the CA monitors the specific memory in the MCU.
If a payload is written, the CA reads the binary data from the specific memory and delivers them to the TA.
Then the TA deserializes the received binary data and executes them one by one to test a Trusted OS implementation.
Accordingly, the TA invokes specific syscalls according to the ordinal numbers and fills them in with arguments extracted from the payload.
To this end, the TA hardcodes the syscalls declaration codes that are manually prepared.
The manual work to prepare the declaration codes is a one-shot effort that can be easily done by referring to the GP TEE Internal Core API specification. 

\subsection{Selective Instruction Trace Tracking}\label{sec:tracetracking}
This section introduces how to obtain the code coverage feedback in \system{} when testing the Trusted OSes.
To handle the inability of instrumentation in Challenge \uppercase\expandafter{\romannumeral1}, we present a selective instruction trace track, which is implemented in the trace collector in the fuzzing engine. 
The trace collector controls the debug probe to collect traces synchronously when the test cases are being executed.
After completing a test case, it calculates the code coverage and delivers it to the manager as feedback.

In the selective instruction trace track, we utilize the ETM component to enable non-invasive monitoring of the execution context of target Trusted OSes.
The overall workflow is as follows.
(1) Before each payload is sent to the MCU, the hardware-assisted controller resets the target MCU via the debug probe.
(2) Once the execution engine reads a valid payload from the specific memory space and starts to execute it, the hardware-assisted controller starts the ETM component to record the instruction trace for each syscall from the payload.
Specifically, in the meanwhile of executing the syscalls, the generated instruction trace is synchronously recorded and delivered to the fuzzing engine via the debug probe.
(3) After completing a payload, the hardware-assisted controller computes the coverage based on the instruction traces for the manager.

\begin{lstlisting}[language=C,caption={A code snippet containing the main execution logic from our designed TA.},label={lst:ta-proxy}][t]
  extern int32_t start_event;
  extern int32_t stop_event;
  void TA_ProcessEachPayload(){
    RecvPayloadFromCA();
    DecodePayload();
    //Start invoking syscalls one by one
    do {
      // Data access event is triggered, start tracing
      start_event++;
      InvokeOneSyscall();
      // Data access event is triggered, stop tracing
      stop_event++;
      if (AllSyscallsExecuted()){
        break;
      }
    } while(1);
  } 
  \end{lstlisting}

However, the IoT Trusted OSes are highly resource-constrained, making locally storing ETM traces infeasible and limiting the speed of local fuzzing.
Therefore, \system{} utilizes the debug probe (see Section~\ref{sec:fuzzingframework}) to stream all ETM trace data to the host PC in real time while the target system is running.
Moreover, \system{} enables parallel execution of test case generation, transmission, and execution, as well as coverage calculation, thereby boosting the speed of fuzzing.

For code coverage, we cannot directly use the raw instruction trace packets generated by the ETM as a replacement for branch coverage due to two issues.
First, there is a gap between the raw ETM trace packets and the instruction traces generated by the Trusted OS.
The TEE internals are complicated and the ETM component records instruction traces generated by the software running on the MCU, including the CA, rich OS, the TA, and the Trusted OS.
Thus, we design a selective instruction trace collection strategy to generate fine-grained traces.
ARM ETM allows enabling/disabling trace collection when corresponding events occur.
We configure the different events via the Data Watchpoint and Trace Unit (DWT) hardware feature to filter out noisy packets.
In \system{}, we aim to calculate the code coverage triggered by every syscall in a test case and generated by Trusted OS.
Thus, as shown in Listing~\ref{lst:ta-proxy}, we configure the event-based filters by adding two data write access event conditions.
The two event conditions start ETM tracing before invoking a syscall and stop ETM tracing after completing the syscall, respectively.
In addition, we specify the address range of the secure world shall be included in the trace stream to filter noisy trace packets generated from the normal world.

Second, there is a gap between the raw ETM trace packets to the quantitive coverage results.
To precisely recover the branch coverage information, we have to decode the raw trace packets and map them to disassembled binary instruction address.
After that, we can recover the instruction traces and construct the branch coverage information\cite{zhou2022ncscope} \cite{xue2021happer}.
However, disassembling code introduces significant run-time overhead as it incurs high computation cost \cite{chen2019ptrix}.
Thus, we calculate the branch coverage directly using the raw trace packets \cite{li2022mu}.
At a high level, this calculation mechanism utilizes a special basic block generated with Linear Code Sequence and Jump (LCSAJ) \cite{yates1995effort} to reflect any change in basic block transitions.
LCSAJ basic blocks consist of the basic address of a raw ETM branch packet and a sequence of branch conditions.
The branch conditions indicate whether the instructions followed by the basic address are executed.
This mechanism performs several hash operations on the LCSAJ basic blocks to transform them into random IDs, which we utilize as branch coverage feedback.

\subsection{State Variable Inference and Monitoring}\label{sec:stateinfer}
Here we introduce how to identify the internal state of the Trusted OS and how to obtain the state coverage feedback through the state variable monitor component.
In particular, the state variable inference provides the state variable monitor with the address ranges of the inferred state variables.
Then the state variable monitor tracks the values of these state variables synchronously when the execution engine executes test cases.
After completing a test case, the state variable monitor calculates the state coverage and delivers it to the manager as feedback.
Below are the details about state variable inference and monitoring.

According to the GP TEE Core API specification, Trusted OSes have to maintain several complex state machines to achieve the cryptographic algorithm in a highly secure way.
To explore all states of Trusted OS, the fuzzer needs to feed syscall sequences in several specific orders with different specific state-related argument values.
Coverage-based fuzzers are unlikely to explore further states, causing the executions to miss some vulnerabilities hidden in a deep state. 
For instance, a syscall sequence achieves a new DES cryptographic operation configuration by filling different arguments, whose code coverage may be the same as the syscall sequence to achieve a new AES cryptographic operation configuration.
Preserving such syscall sequences as a seed and further exploring them will achieve new cryptographic operations and gain new code coverages. 
However, a coverage-based fuzzer may discard such syscall sequences that seem to have no code coverage contribution but trigger new internal states.
Thus, \system{} additionally adopts state coverage as feedback to handle the statefulness of Trusted OSes.

\textbf{State variable inference.} By referring to the GP TEE Internal Core API specification and several open-source TEE implementations from Github, we find that Trusted OSes maintain two important structures, including the \texttt{struct} named \texttt{TEE\_OperationHandle} and the \texttt{struct} named \texttt{TEE\_ObjectHandle}, which present the internal states and control the execution context.
We further find that several vital variables (with names such as \texttt{operationState} and \texttt{flags}) in the two complex \texttt{structs} determine the Trusted OS' internal state.
Thus, we utilize the value combinations of state variables to present the states of Trusted OS and track all state-related variables to collect their values.
\iffalse
Then we consider a syscall triggering a new value of the combination of all state variables as triggering a new state.
\fi
Then, we consider a new value of the state variable combination as a new state.
However, the \texttt{TEE\_OperationHandle} and \texttt{TEE\_ObjectHandle} implementations are customized and close source, making recognizing the state variables and their addresses challenging.

To handle it, we come up with an active inference method to recognize the state variables in the \texttt{TEE\_OperationHandle} and \texttt{TEE\_ObjectHandle} implementations.
This method is based on the assumption that the state variables in the above two handles will have different values according to different cryptographic operation configurations.
For instance, in Samsung's Trusted OS implementation, after executing several syscalls to set cryptographic arguments, the value of a state variable from  \texttt{TEE\_OperationHandle} changes from 0 to 1, which means a cryptographic operation is initialized.

\begin{figure}[tbp]
  \centering
  \includegraphics[width=0.45\textwidth]{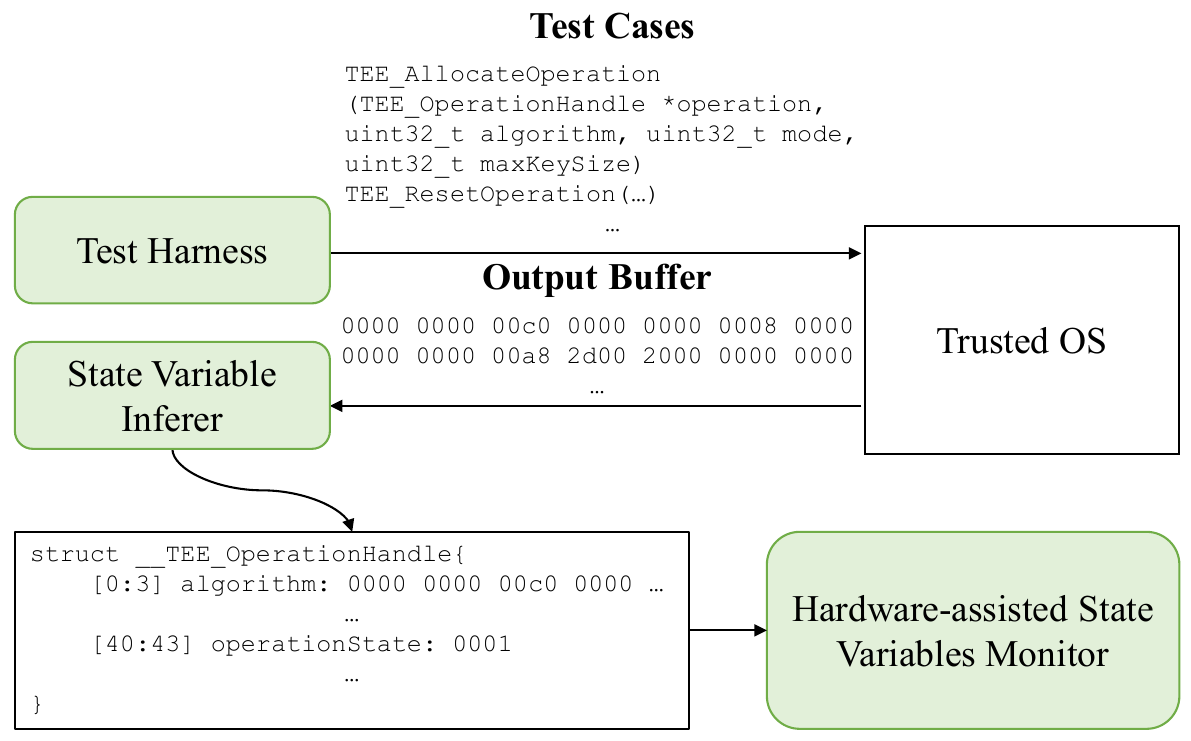} 
  \caption{State variable inference. }
  \label{fig:stateinfer} 
\end{figure}

Based on the assumption, \system{} uses a test harness to generate and execute test cases carefully. 
Then \system{} records the buffers of the two handles and applies state variable inference to detect the address ranges of state variables in the recorded buffers, as shown in Figure~\ref{fig:stateinfer}.
\system{} first filters randomly changeable byte sequences that store pointers, encryption keys, and cipher text.
Specifically, \system{} conducts a 24-hour basic fuzzing procedure on the initial seeds from Section~\ref{sec:input} and collects the buffers of the two handles.
\system{} then parses the buffer into four-byte sequences to recognize changeable values.
After that, \system{} collects all different values for each byte sequence in the fuzzing procedure and calculates the number of times that these different values occur.
\system{} considers the byte sequences that occur over 80 times as buffers and excludes them from the state variables.
For the remaining byte sequences, \system{} applies the following inference to identify state variables.
In our observations, the cryptographic operation configurations are determined by the operation-related arguments, including the operation mode (encryption, decryption, or verification) and cryptographic algorithm (AES, DES, or MAC).
The cryptographic operation configurations are also determined by the operation-related syscall sequences.
We identify such arguments and syscalls by referring to GP TEE Internal Core API specification and conclude the operation-related syscalls include the syscalls that accept the two handles as an input argument and are specified to allocate, init, update, and finish a cryptographic operation, such as \texttt{TEE\_AllocateOperation} and \texttt{TEE\_AllocateTriansientObject}.
\system{} performs mutated seeds that include the operation-related syscalls and records the buffers of the two handles.
These byte sequences that vary with certain syscall sequences with certain arguments are considered state variables.
Finally, \system{} outputs the address ranges of these byte sequences and feeds them to the hardware-assisted controller.

\textbf{State variable monitor.}
The identified state variables and their address ranges are then used as configurations of the hardware-assisted controller. 
\system{} utilizes the debug probe to monitor the state variables.
When a new \texttt{TEE\_OperationHandle} or \texttt{TEE\_ObjectHandle} is allocated, \system{} records its start address.
Given the start address of the two handles and address ranges of state variables, \system{} calculates the memory ranges of state variables and directly reads the memory via the debug probe.
In the fuzzing loop, for each syscall, beside the branch coverage, \system{} records the hash of combination values of the state variables as the state coverage.

\subsection{Fuzzing Loop}\label{sec:fuzzingloop}
After collecting code and state coverage feedback, the fuzzer enters the main fuzzing loop implemented in the manager.
\system{} has a similar basic fuzzing loop to \syzkaller{}.
It schedules two tasks for test case generation:  generating a new test case from syscall templates or mutating an existing one based on selected seeds. 
As for the generation task, \system{} faithfully borrows the generation strategy from \syzkaller{}.
As for the mutation task, \system{} utilizes a composite feedback mechanism to explore rarely visited states while increasing code coverage.
Notably, \system{} does not adopt the \textbf{Triage} scheduling tasks from \syzkaller{} due to two key reasons.
First,  \system{} fuzzes directly on the MCU, enabling swift MCU resets after each test case, thus mitigating false positives in branch coverage.
Branch coverage validation through triaging is, therefore, unnecessary.
Second, executing test cases for Trusted OSes (averaging 30 syscalls) is time-consuming, incurring a significant overhead for \system{} to minimize a syscall sequence by removing syscalls one by one.
Appendix~\ref{app:motivation} further experimentally demonstrates our intuition for the new scheduling tasks.

\subsection{Composite Feedback Mechanism}\label{sec:feedback}
\vspace{-2mm}
\begin{figure}[tbp]
  \centering
  \includegraphics[width=0.40\textwidth]{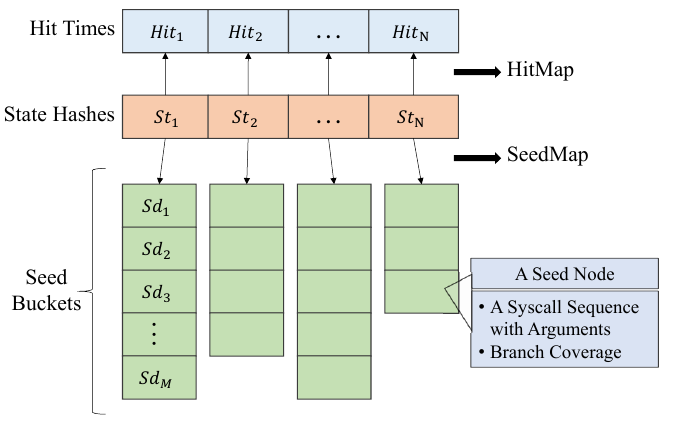} 
  \caption{Seed corpus in \system{}.}
  \label{fig:seedcorpus} 
\end{figure}
\system{} adopts a novel composite feedback mechanism, leveraging both code and state coverage to guide mutation tasks within the fuzzing loop.
Specifically, \system{} preserves and prioritizes seeds that trigger new code or states.
We design two maps as the seed corpus to preserve seeds according to code coverage and state feedback.
With such seed corpus, \system{} then periodically selects seeds from the hash table to explore new codes and new states.
This section introduces how \system{} fine-tunes the evolution through a novel composite feedback mechanism, including seed preservation and seed selection strategy.

\textbf{Seed preservation.}
Given the composite feedback mechanism, we thus provide two maps as the seed corpus to store seeds that discover new state values and new code, respectively.
As shown in Figure~\ref{fig:seedcorpus}, in two maps, \system{} calculates the hash of combination values of state variables as the keys.
\system{} maps the state hashes to their hit times in the HitMap and maps the state hashes to seed buckets in the SeedMap.
In the SeedMap, for a certain state hash, the mapping seed bucket contains one or several seeds that can produce the matching state variable values.
To construct the SeedMap, \system{} handles the following two situations.
First, if a syscall from a test case triggers a new value combination of state variables, \system{} adds a new state hash in the SeedMap.
Then, \system{} maps the new state hash to a new seed bucket.
To construct the seed bucket, \system{} utilizes the test case with its branch coverage to construct a seed node and appends the seed bucket with the newly constructed seed node.
Second, if a syscall from a test case triggers a new code coverage, \system{} adds a new seed node in a seed bucket.
Specifically, \system{} calculates the hash of combination values of state variables produced by the syscall.
Then, \system{} looks up the seed bucket that is mapped to the state hash and appends a new seed node that contains the test case and its branch coverage to the seed bucket.
Noted, a test case could be stored in multiple seed buckets if it triggers multiple feedbacks at the same time.
As for the HitMap, each time after completing a test case, \system{} calculates all the state hashes it triggers and updates the hit times for these state hashes according to their hit times.

\begin{algorithm}[t]
  \SetAlgoLined
  \caption{Seed Selection Algorithm}
  \label{algo:seed-selection-algorithm}
  \KwIn{$C$, SeedMap that maps state hashes to a set of seeds}
  \KwIn{$H$, HitMap that maps state hashes to hit times}
  \KwOut{$s$, the selected seed}
  $Sum\_W\gets 0$\; 
  $Map\_W=$ initMap() //maps state hashes to their weights \;
  \For {$each\ key \in H$}{
    $t\gets$ getValue($H, key$)\;
    $Sum\_W\gets Sum\_W + t^{-1}$\;
  }
  \For {$each\ key \in H$}{
    $t\gets$ getValue($H,key$)\;
    $Map\_W\gets Map\_W\cup (key,t^{-1}/Sum\_W)$\;
  }
  $St\gets$ WeightedRandom\_StateSelection($Map\_W$)\;
  $seedSets\gets$ getValue($C,St$)\;
  $s\gets$ WeightedRandom\_SeedSelection($seedSets$)\;
  \end{algorithm}
  
\textbf{Seed selection.}
Given the preserved seed corpus, \system{} applies a special seed selection strategy to improve the fuzzing efficiency.
Algorithm~\ref{algo:seed-selection-algorithm} shows how \system{} selects seeds from the corpus.
First, \system{} chooses a state and then chooses a seed from the mapping seed buckets according to the state.
In state selection, \system{} is more likely to choose a rarely visited state by a weighted random selection algorithm.
The probability of choosing a seed is negatively correlated with its hit times.
Thus, we assign each seed a weight value, which is the reciprocal of its hit times.
Finally, The probability of choosing a seed is equal to the proportion of its weight in the sum of all weights.
In seed selection, \system{} is more likely to choose a seed with high branch coverage.
To this end, \system{} chooses a seed based on a weighted random selection algorithm, and the probability of choosing a certain seed is equal to the proportion of its branch coverage in all coverages.
Notably, the probabilities of choosing a state and seed are dynamically updated since the hit times and the sum of coverage is updated in the fuzzing procedure.

\subsection{Scope and Scalability of \system{}.}\label{sec:scopeandscalability}
\system{} targets Trusted OSes provided by IoT vendors and assumes that (i) a TA can be installed in the Trusted OS, and (ii) target devices have ETM enabled.
These assumptions align with typical IoT Trusted OS scenarios.
First, given that IoT device manufacturers often need to implement device-specific TAs, Trusted OS binaries supplied by IoT vendors generally allow TA installation. Second, \system{} tests IoT Trusted OSes by deploying them on development boards where ETM is enabled by default.
Moreover, \system{} can directly test the Trusted OSes following the GP TEE Internal Core API specification with MCU-specific configurations.
It has built-in support for testing on alternative Trusted OSes, including proprietary ones.
Appendix~\ref{app:assumption} extends discussion with concrete data.

\system{} can support other Trusted OSes.
To test a new Trusted OSes on a different MCU, \system{} requires MCU configurations, including the address ranges of specific memory for storing payloads, the addresses of the data events for the event-based filter, and the address ranges of secure memory for the address-based filter.
We developed tooling in the CA and TA to automatically help the analyst obtain all required addresses.
In addition, by following the development documents from IoT TEE vendors, the CA and TA may require slight adjustments to meet the format required by the new Trusted OSes and loaded into the Rich OS and Trusted OS.

To extend \system{} to proprietary Trusted OSes, we augment the syscall templates and the API declarations in our designed TA and test harness with the new version of these customized APIs.
This can be done by referring to the API documents provided by IoT vendors, which is simple and requires minimum effort.
To enable the state-aware feature, we need expert analysis of the state-related structures in the Trusted OS and the use of our state variable inference to collect address ranges. 
These structures can be extracted from the documents and header files. 
We rely on two heuristics to help extract them. First, state-related data structures usually have common names, e.g., related to $context$ or $state$. 
Second, the state structures will be the inputs and outputs of several crypto operation-related syscalls. 
For example, on Link TEE Air, a pointer
named $context$ is used among cryptographic syscalls such as $tee\_aes\_init$ and $tee\_aes\_update$, and can be further utilized to infer state variables. 
This information can be obtained from the $crypto.h$ header file.

%%%%%%%%%%%%%%%%%%%%%%%%%%%%%%%%%%%%%%%%%%%%%%%%%%%%%%%%%%%%%%%%%%%%%%%%%%%%%%%%
\section{Implementation}
\label{sec:impl}
%%%%%%%%%%%%%%%%%%%%%%%%%%%%%%%%%%%%%%%%%%%%%%%%%%%%%%%%%%%%%%%%%%%%%%%%%%%%%%%%
We have implemented a prototype of \system{} on top of \syzkaller{}.
We replaced \syzkaller{}'s execution engine with our custom CA and TA pair, integrating our extended syscall templates, eliminating \syzkaller{}'s triage scheduling task, and implementing our own seed preservation and selection strategy. Sections~\ref{sec:fuzzingframework}, \ref{sec:fuzzingloop}, and \ref{sec:feedback} detail these adaptations.

\begin{figure}[tbp]
  \centering
  \includegraphics[width=0.35\textwidth]{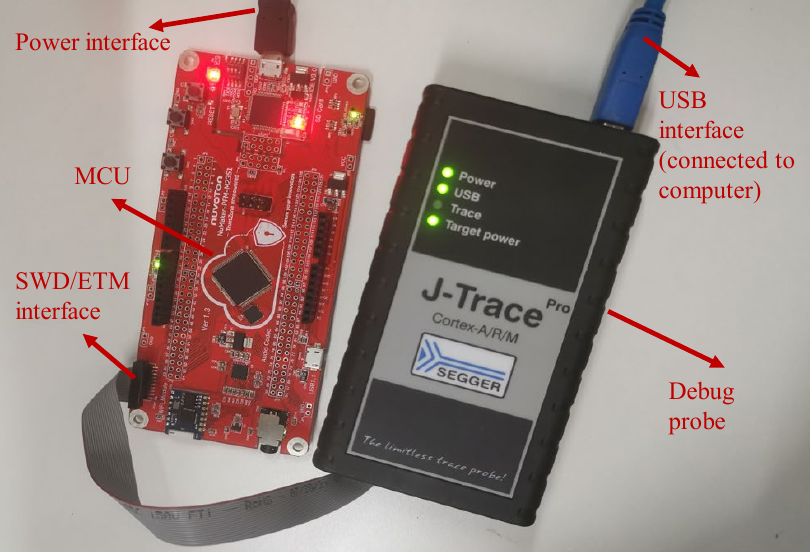} 
  \caption{\system{} setup in fuzzing a target TEE implementation on the Nuvoton M2351 board. 
  The debug probe accesses memory and tracks instruction traces on the MCU via the SWD/ETM interface. 
  It also delivers data to the PC and receives commands from the PC via a USB interface.
  }
  \label{fig:probe} 
\end{figure}

Below are details about our implementations.
(1) As for the overall fuzzing framework, we use the SEGGER J-Trace Pro debug probe to control the communication between the fuzzing engine and the execution engine, as shown in Figure~\ref{fig:probe}.
The pair of CA and TA is developed following the GP TEE Internal Core API specification and is loaded into the MCU following the instructions provided by the IoT vendors.
To control the debug probe, we developed a hardware-assisted controller based on the SEGGER J-Link SDK.
The hardware-assisted controller receives commands from the manager and sends feedback collected on the MCU to the manager via socket communications.
(2) For the selective instruction trace track, \system{} integrates the ETM tracing component of the SEGGER J-Trace Pro and records the instruction traces from Trusted OSes non-invasively.
The raw ETM packets decoder and branch coverage calculation are accomplished in the hardware-assisted controller.
(3) For the state variable inference and monitoring, \system{} follows the testing strategy in Section~\ref{sec:stateinfer} and utilizes an RTT component \cite{rtt} from the SEGGER J-Trace Pro to record related state variable values and deliver them to the fuzzing engine.
The RTT component accesses memory in the background with high-speed transmission.

Several tools help us analyze the root cause of detected crashes.
We utilize CmBacktrace, a customized backtrace tool \cite{cmbacktrace}, to track and locate error codes automatically.
Additionally, we develop TEEKASAN based on KASAN \cite{ksan} and MCUASAN \cite{mcuasan} to help identify out-of-bound and use-after-free vulnerabilities.
We integrate TEEKASAN with lightweight compiler instrumentation and develop shadow memory checking for bug triaging on the open source Trusted OSes.
Since TEEKASAN only analyzed a small number of vulnerabilities due to the limitation of the instrumentation tool and most Trusted OSes are closed-source, we manually triage the remaining vulnerabilities.

%%%%%%%%%%%%%%%%%%%%%%%%%%%%%%%%%%%%%%%%%%%%%%%%%%%%%%%%%%%%%%%%%%%%%%%%%%%%%%%%
\section{Evaluation}
%%%%%%%%%%%%%%%%%%%%%%%%%%%%%%%%%%%%%%%%%%%%%%%%%%%%%%%%%%%%%%%%%%%%%%%%%%%%%%%%
In this section, we comprehensively evaluate the \system{}, demonstrating its effectiveness in fuzzing IoT Trusted OSes.
First,  we evaluate the overhead breakdown of \system{}.
Second, we conduct experiments exploring the effectiveness of our designs.
Third, we examine \system{}'s state variable inference capabilities.
Finally, we apply \system{} on fuzzing three real-world IoT Trusted OSes and introduce their vulnerabilities.
In summary, we aim to answer the following research questions:

\noindent \textbf{RQ1:} What is the overhead breakdown of \system? (Section~\ref{sec:overhead})

\noindent \textbf{RQ2:} Is \system effective for fuzzing IoT Trusted OSes? (Section~\ref{sec:abulation})

\noindent \textbf{RQ3:} Is the state variable inference method effective, and are the inferred state variables expressive? (Section~\ref{sec:evaluatestate})

\noindent \textbf{RQ4:} How vulnerable are the real-world Trusted OSes from different IoT vendors from \system's results? (Section~\ref{sec:overallperformance})

\subsection{Experimental Setup}

\noindent\textbf{Target Trusted OSes.}
We evaluate \system{} on three Trusted OSes designed for IoT devices, mTower, TinyTEE, and Link TEE Air.
The reason for selecting these targets is as follows.
mTower and TinyTEE both provide the standard APIs following the GP TEE Internal Core API specification. 
In addition, they are developed by two leading IoT vendors, Samsung and Tsinglink Cloud (which serve more than 30 downstream IoT manufacturers, including China TELECON and Panasonic), respectively.
Link TEE Air is a proprietary Trusted OSes developed by Ali Cloud for \system{} to evaluate its built-in support of closed-source and proprietary Trusted OSes.
Moreover, the three targets have been adopted in a number of IoT devices and their security vulnerabilities have a practical impact.
In this paper, we evaluate mTower v0.3.0, TinyTEE v3.0.1, Link TEE Air v2.0.0, and each of them is the latest version during our experiments.

\noindent\textbf{Experiment settings.}
We perform our evaluation under the same experiment settings: a personal computer with 3.20GHz i7-8700 CPU, 32GB RAM, Python 3.8.2, Go version 1.14, and Windows 10.

\noindent\textbf{Evaluation metrics.}
We evaluate \system{} with the following three aspects.
(1) We measure the branch coverage to evaluate the capability of exploring codes, which is widely used in recent research \cite{buschteezz}\cite{harrison2020partemu}.
The branch calculation is introduced in Section~\ref{sec:tracetracking}.
(2) To evaluate the capability of exploring the deep state, we measure the state coverage and the syscall sequence length of each fuzzer.
Specifically, we consider the value combinations of state variables as a state and measure the number of different states.
(3) We count the number of unique vulnerabilities.
\system relies on the built-in exception handling mechanism to detect abnormal behaviors of Trusted OSes.
We explore the dedicated fault status registers \cite{faulthandling} to identify the \texttt{HardFault} exception in concerns.
These exceptions indicate critical system errors and thus can be used as a crash signal \cite{CortexM23}.
For the crashes, we reproduce them and report their stack traces by CmBackTrace to track and locate error codes.
We filter stack traces into unique function call sequences to collect the explored unique bugs on target programs, which are widely used for deduplication in the CVE dataset \cite{cve} and debugging for vendors.
Following best practices, we extract the top three function calls in the stack traces to de-duplicate bugs \cite{klees2018evaluating,li2021unifuzz}.
We then analyze the root cause of the bugs manually. 

\noindent\textbf{State transition analysis.}
We further develop a script to automatically construct a state transition tree, which helps visually understand the practical meaning of the state variables.
Specifically, we utilize the state hashes calculated based on state variables to present the states of Trusted OSes and take the syscall sequences as state transition labels.

\begin{figure}[tbp]
  \centering
  \includegraphics[width=0.4\textwidth]{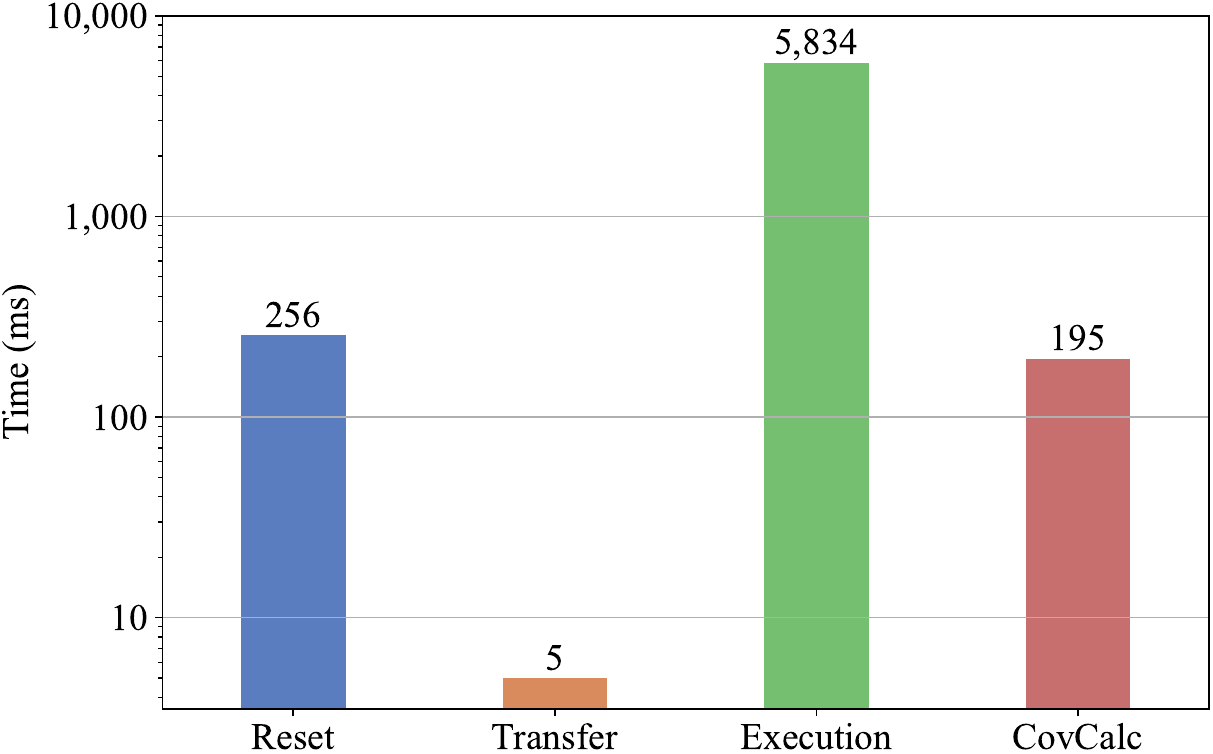} 
  \caption{Overhead breakdown of \system. }
  \label{fig:overhead-breakdown} 
\end{figure}

\subsection{Overhead Breakdown (RQ1)}\label{sec:overhead}
We measure the execution time of sub-processes of \system to assess its impact on the overall overhead.
For each fuzzing round, \system performs the following sub-processes: 
(1) \texttt{Reset} means the time spent on the MCU resetting.
(2) \texttt{Transfer} means the time spent on the manager engine sending the test case.
(3) \texttt{Execution} means the time spent on the execution engine executing the test case. 
In the meanwhile, the debug probe records the raw instruction trace packets and state variable values and transfers them to the PC via RTT.
(4) \texttt{CovCalc} means the time spent on the hardware-assisted controller decoding the collected raw trace packets and calculating the branch coverage.
In the meanwhile, the hardware-assisted controller calculates the state hashes based on the state variable values.
We only evaluate the sub-processes that are related to interacting with the resource-limited MCU, which mainly determines the speed of \system{}.
As introduced in Section~\ref{sec:tracetracking}, the subprocesses of \texttt{Transfer}, \texttt{Execution}, and \texttt{CovCalc} are designed to run in parallel. 

The results are shown in Figure~\ref{fig:overhead-breakdown}, from which we have the following conclusions.
First, the overheads of \texttt{Reset}, \texttt{Transfer}, and \texttt{CovCalc} in \system are relatively low.
Thus, using \texttt{Reset} to mitigate the false positive coverage and vulnerabilities is acceptable.
Second, \texttt{Execution} takes the most time.
Test cases for IoT Trusted OSes include 14.4 syscalls on average and are complex.
In addition, for \texttt{Execution}, the Nuvoton M2351 board used in our manuscript has no Embedded Trace Buffer (ETB), \system{} utilizes the debug probe to stream all the ETM trace data to the host PC in real time.
We conducted a 48-hour fuzzing and found the peak tracing speed is \etmtracingspeed{}/s, and 92\% of the trace files for single syscalls are smaller than 2KB.
In conclusion, the ETM tracing takes less time than the syscall execution.
\vspace{-1mm}
\begin{mybox}
  \textbf{RQ1:} 
  It takes \system{} 6,290 ms on average to complete a test case and to collect its feedback.
  The subprocess of executing a test case on the MCU takes the most time, while the orchestration and analysis take only roughly 1\% of the overall time.
\end{mybox}

\subsection{Effectiveness of \system{} (RQ2)}\label{sec:abulation}
\begin{figure*}[t]
\setlength{\abovecaptionskip}{0.3cm}
    \setlength{\belowcaptionskip}{-0.3cm}
      \begin{subfigure}[b]{0.24\textwidth}    
          \centering
          \includegraphics[width=1.71in,height=1.30in]{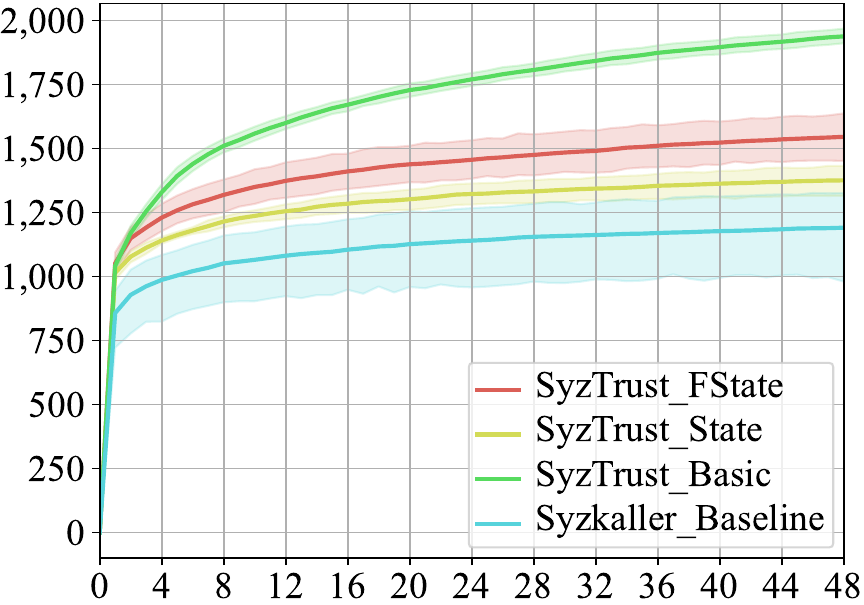}
          \caption{Branch coverage.}
          \label{fig:codecov}
      \end{subfigure}
         \begin{subfigure}[b]{0.24\textwidth}
             \centering
             \includegraphics[width=1.71in,height=1.30in]{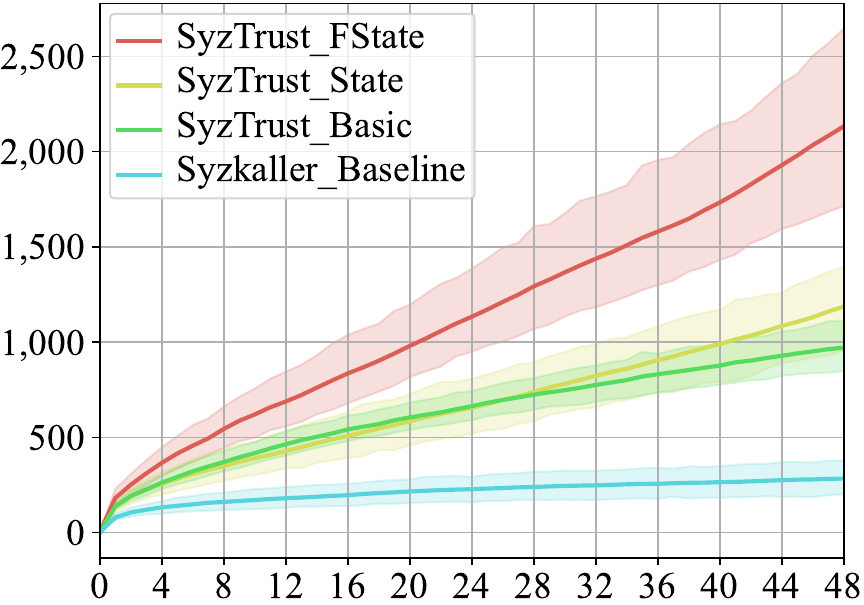}
             \caption{States (based on variables).}
             \label{fig:statev}
         \end{subfigure}
         \begin{subfigure}[b]{0.24\textwidth}
          \centering
          \includegraphics[width=1.71in,height=1.30in]{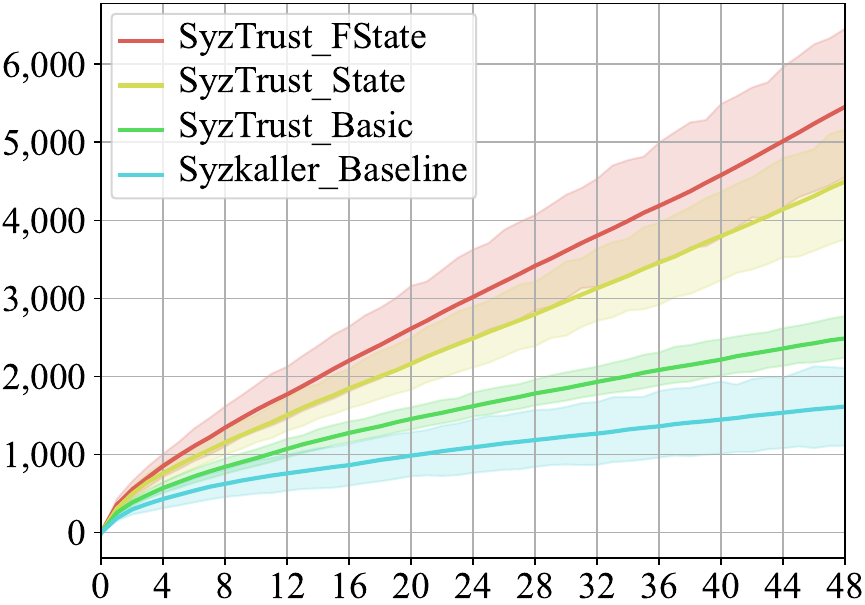}
          \caption{States (based on handles).}
          \label{fig:stateh}
      \end{subfigure}
         \begin{subfigure}[b]{0.24\textwidth}
             \centering
             \includegraphics[width=1.71in,height=1.30in]{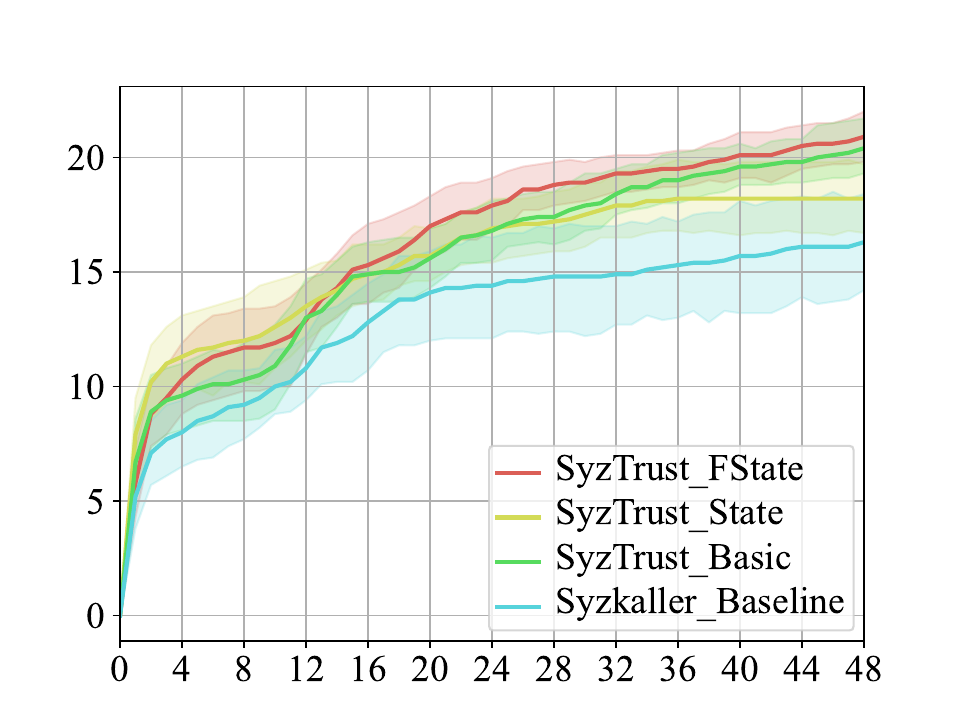}
             \caption{Unique vulnerabilities.}
             \label{fig:uniquebugs}
         \end{subfigure}
             \caption{The branch coverage growth, state numbers growth, and unique vulnerability growth discovered by \syzkaller{}, \sysbasic{}, \sysstate{}, and \sysfstate{}. 
             \iffalse
             where \sysbasic only implements the new scheduling tasks and extends the syscall templates with helper syscalls, \sysstate$\ $implements the composite feedback mechanism in considering the complete values of two handles as state variables, and \sysfstate implements composite feedback mechanism with inferred fine-grained state variables.
             \fi
             } 
             \label{fig:abulationresult}     
 \end{figure*}

In this section, we evaluate \syzkaller{}, \sysbasic{}, \sysstate{}, and \sysfstate{} on mTower to explore the effectiveness of our designs, with each experiment running for 48 hours.
To measure our new scheduling tasks and composite feedback mechanism, we construct two prototypes of \system only with the new scheduling tasks and only with the composite feedback mechanism, which are named \sysbasic and \sysfstate, respectively.
In addition, to measure the necessity of our state variable inference, we construct a prototype that considers the complete buffer values of two state handles as state variables named \sysstate.
We evaluate three prototypes of \system against the state-of-the-art fuzzer \syzkaller, which has found a large number of vulnerabilities on several kernels and is actively maintained.
To reduce the randomness, we repeat all experiments ten times. 
The results are shown in Figure~\ref{fig:abulationresult} and Table~\ref{tab:allbugs}.

\begin{table}[tbp]
  \centering
  \caption{The unique vulnerabilities found by \syzkaller{}, \sysbasic{}, \sysstate{}, and \sysfstate{} in 48 hours. The vulnerabilities whose IDs are marked with * have already received CVEs.}
  \resizebox{0.80\linewidth}{!}{
  \begin{tabular}{@{}ccccc@{}}
    \toprule
    Vul. ID         & Syzkaller-Baseline & SyzTrust-Basic & SyzTrust-State & SyzTrust-FState \\ \midrule
    1*              & \fullcirc          & \fullcirc      & \fullcirc      & \fullcirc       \\
    2*             & \fullcirc          & \fullcirc      & \fullcirc      & \fullcirc       \\
    3              & \emptcirc          & \fullcirc      & \emptcirc      & \fullcirc       \\
    4              & \emptcirc          & \emptcirc      & \fullcirc      & \emptcirc       \\
    5*              & \fullcirc          & \fullcirc      & \fullcirc      & \fullcirc       \\
    6              & \emptcirc          & \emptcirc      & \fullcirc      & \emptcirc       \\
    7*              & \emptcirc          & \emptcirc      & \fullcirc      & \fullcirc       \\
    8*            & \fullcirc          & \fullcirc      & \fullcirc      & \fullcirc       \\
    9*             & \fullcirc          & \fullcirc      & \fullcirc      & \fullcirc       \\
    10              & \emptcirc          & \emptcirc      & \fullcirc      & \fullcirc       \\
    11             & \fullcirc          & \emptcirc      & \emptcirc      & \fullcirc       \\
    12*             & \fullcirc          & \emptcirc      & \fullcirc      & \fullcirc       \\
    13*            & \fullcirc          & \fullcirc      & \fullcirc      & \fullcirc       \\
    14*           & \fullcirc          & \fullcirc      & \fullcirc      & \fullcirc       \\
    15*            & \fullcirc          & \fullcirc      & \fullcirc      & \fullcirc       \\
    16             & \fullcirc          & \fullcirc      & \fullcirc      & \fullcirc       \\
    17             & \emptcirc          & \emptcirc      & \emptcirc      & \fullcirc       \\
    18             & \emptcirc          & \emptcirc      & \emptcirc      & \fullcirc       \\
    19             & \emptcirc          & \emptcirc      & \emptcirc      & \fullcirc       \\
    20             & \fullcirc          & \fullcirc      & \fullcirc      & \fullcirc       \\
    21             & \fullcirc          & \fullcirc      & \fullcirc      & \fullcirc       \\
    22             & \fullcirc          & \fullcirc      & \fullcirc      & \fullcirc       \\
    23             & \fullcirc          & \fullcirc      & \fullcirc      & \fullcirc       \\
    24             & \fullcirc          & \fullcirc      & \fullcirc      & \fullcirc       \\
    25             & \fullcirc          & \fullcirc      & \fullcirc      & \fullcirc       \\
    26             & \fullcirc          & \fullcirc      & \fullcirc      & \fullcirc       \\
    27             & \fullcirc          & \fullcirc      & \fullcirc      & \fullcirc       \\
    28             & \fullcirc          & \fullcirc      & \fullcirc      & \fullcirc       \\
    29             & \fullcirc          & \fullcirc      & \fullcirc      & \fullcirc       \\
    30             & \fullcirc          & \fullcirc      & \fullcirc      & \fullcirc       \\
    31             & \fullcirc          & \fullcirc      & \fullcirc      & \fullcirc       \\
    32             & \fullcirc          & \fullcirc      & \fullcirc      & \fullcirc       \\
    33             & \emptcirc          & \fullcirc      & \emptcirc      & \fullcirc       \\
    34             & \emptcirc          & \fullcirc      & \fullcirc      & \fullcirc       \\
    35             & \fullcirc          & \fullcirc      & \fullcirc      & \fullcirc       \\
    36             & \emptcirc          & \emptcirc      & \fullcirc      & \fullcirc       \\
    37             & \emptcirc          & \fullcirc      & \fullcirc      & \fullcirc       \\
    38             & \fullcirc          & \fullcirc      & \emptcirc      & \emptcirc       \\ \midrule
    \textbf{Total} & \textbf{26}        & \textbf{28}    & \textbf{31}    & \textbf{35}     \\ \bottomrule
    \end{tabular}
  }
  \label{tab:allbugs}
  \end{table}

Branch coverage is calculated in terms of LCSAJ basic block number introduced in  Section~\ref{sec:tracetracking}. \syzkaller{} triggers \syzkallercode{} branches on average and is exceeded by all three versions of \system shown in Figure~\ref{fig:abulationresult}. 
Among them, \sysbasic{} archives the highest code coverage among the four fuzzers with \syzbasiccode{} branches, which shows the effectiveness of our new scheduling tasks and the extended syscall templates.
When state coverage feedback is integrated into \system, the branch coverage explored by the \sysstate{} and \sysfstate{} is lower than \sysbasic{}.
It is because their adopted composite feedback mechanisms drive them to preserve more seeds that trigger new states, which might have no contribution to the code coverage.
Given the same evaluation time, a certain percentage of time is assigned to mutating and testing such seeds that trigger new states, and the code coverage growth will result in slower growth.
Thus, we additionally perform a long-time fuzzing experiment, and the result shows that \sysfstate{} can achieve 1,984 branches in 74 hours.

Among the four fuzzers, \sysfstate{} triggers the most states with 2,132 states on average, and \syzkaller{} triggers the least states with 284 states on average shown in Figure~\ref{fig:statev}.
\sysstate{} has similar performances on the state growth with \sysbasic{}.
It is because \sysstate{} spends lots of time exploring the test cases that trigger new values of non-state variables since \sysstate{} utilizes the two whole handles' values to present state.
To further illustrate this deduction, we present Figure~\ref{fig:stateh}, where we calculate the growth of the states based on the whole handle values.
In such settings, \sysstate{} triggers more states than \sysbasic{} and \syzkaller{}.
However, these states are not expressive and effective for guiding the fuzzer to trigger more branch coverage and vulnerabilities.

The exploration space of unique vulnerabilities triggered by \syzkaller with \syzkallerbug{} vulnerabilities on average is fully covered and exceeded by the three versions of \system shown in Figure~\ref{fig:abulationresult}.
Among them, \sysfstate detects \fstatebug{} vulnerabilities on average, which achieves the best vulnerability-finding capability.
\sysstate{} has similar performances on vulnerability detection with \sysbasic{} and they detect \syzbasicbug{} vulnerabilities on average.

In addition, Table~\ref{tab:allbugs} shows the unique vulnerabilities detected by \syzkaller and the three versions of \system in ten trials during 48 hours.
The vulnerability ID in Table~\ref{tab:allbugs} is consistent with Table~\ref{tab:bugslist} in Appendix~\ref{app:litedbugs}.
First, \system{} finds all the unique vulnerabilities that \syzkaller finds.
Using the currently assigned CVE as ground truth, \system detected more CVEs than \syzkaller.
Second, \sysfstate finds the most vulnerabilities and finds eight vulnerabilities that \syzkaller and \sysbasic cannot find.
The eight vulnerabilities are all triggered by syscall sequences whose lengths are more than 10, which indicates they are triggered in a deep state.
For instance, the vulnerability of ID 7 occurs when the Trusted OS enters the $key\_set\&initialized$ state after a MAC function is configured and the function $TEE\_MACUpdate$ is invoked with an excessive size value of "chunkSize".
\sysfstate can detect more vulnerabilities, which is benefited from our composite feedback mechanism.
Specifically, the fuzzer preserves the seeds that trigger new states and then can detect more vulnerabilities by exploring these seeds.
In summary, this evaluation reveals two observations.
First, although coverage-based fuzzer achieves high coverage effectively, their vulnerability detection capability will be limited when testing stateful systems. 
Second, for stateful systems, understanding their internal states and utilizing the state feedback to guide fuzzing will be effective in finding more vulnerabilities.

\begin{mybox}
  \textbf{RQ2:} 
  The design of \system{} is effective as the three versions of \system{} outperform \syzkaller{} in terms of code and state coverage and number of detected vulnerabilities.
  New task scheduling with extended syscall templates significantly improves the fuzzer's code exploration, and the composite feedback mechanism helps trigger more states and detect more vulnerabilities.
\end{mybox}

\subsection{State Variable Inference (RQ3)}\label{sec:evaluatestate}
As for state variable inference evaluation, we first evaluate the precision of our state variable inference method.
Second, we utilize the executed syscall sequences and their state hashes to construct a state transition tree and present an example to show the expressiveness of our state variables.

\begin{table}[tbp]
  \centering
  \caption{The number of state variables inferred by \system. False postive is denoted as FP.} 
  \resizebox{0.9\linewidth}{!}{%
  \begin{tabular}{@{}ccccc@{}}
    \toprule
    Target                   & Handle                 & Number & FP & Precision               \\ \midrule
    \multirow{2}{*}{mTower}  & $TEE\_ObjectHandle$    & 11     & 1  & \multirow{2}{*}{87.5\%} \\ \cmidrule(lr){2-4}
                             & $TEE\_OperationHandle$ & 13     & 2  &                         \\ \midrule
    \multirow{2}{*}{TinyTEE} & $TEE\_ObjectHandle$    & 13     & 3  & \multirow{2}{*}{82.6\%} \\ \cmidrule(lr){2-4}
                             & $TEE\_OperationHandle$ & 10     & 1  &                         \\ 
                             \midrule
    \multirow{2}{*}{OP-TEE}  & $TEE\_ObjectHandle$    & 10     & 1  & \multirow{2}{*}{87.0\%} \\ \cmidrule(lr){2-4}
                             & $TEE\_OperationHandle$ & 13     & 2  &                         \\ \midrule
    \multirow{2}{*}{Link TEE Air}  & $context(AES)$    & 6     & 2  & \multirow{2}{*}{71.4\%} \\ \cmidrule(lr){2-4}
    & $context(Hash)$ & 8    & 2  &                         \\ \bottomrule
    \end{tabular}
  }
  \label{tab:stinfer}
  \end{table}

For the precision evaluation, we manually analyze the usage of the inferred state variables in the Trusted OSes.
We check if a state variable is used in condition statements to control the execution context.
As for mTower and OP-TEE, we obtain their source codes and manually read the state variable-related codes.
As for TinyTEE and Link TEE Air, we invite five experts with software reverse engineering experiences to manually analyze their binary codes.

Table~\ref{tab:stinfer} shows the results.
For Trusted OSes, including a proprietary one, our active state variable inference method is effective and achieves \stateinfeacc{}\% precision on average.
These validated state variables are expressive and meaningful, including \texttt{algorithm}, \texttt{operationClass} (description identifier of operation types, e.g., CIPHER, MAC), \texttt{mode} (description identifier of operation, e.g., ENCRYPT, SIGN), and \texttt{handleState} (describing the current state of the operation, e.g., an operation key has been set).
The false positive state variables are of two types.
One is some variables that indicate the length of several specific buffers, e.g., digest length, and have specific values.
Another type is several buffers that do not likely have changeable values. 
Both of them generate a few false positive new states and have little impact on the fuzzing procedure.

\begin{figure}[tbp]
  \centering
  \includegraphics[width=0.4\textwidth]{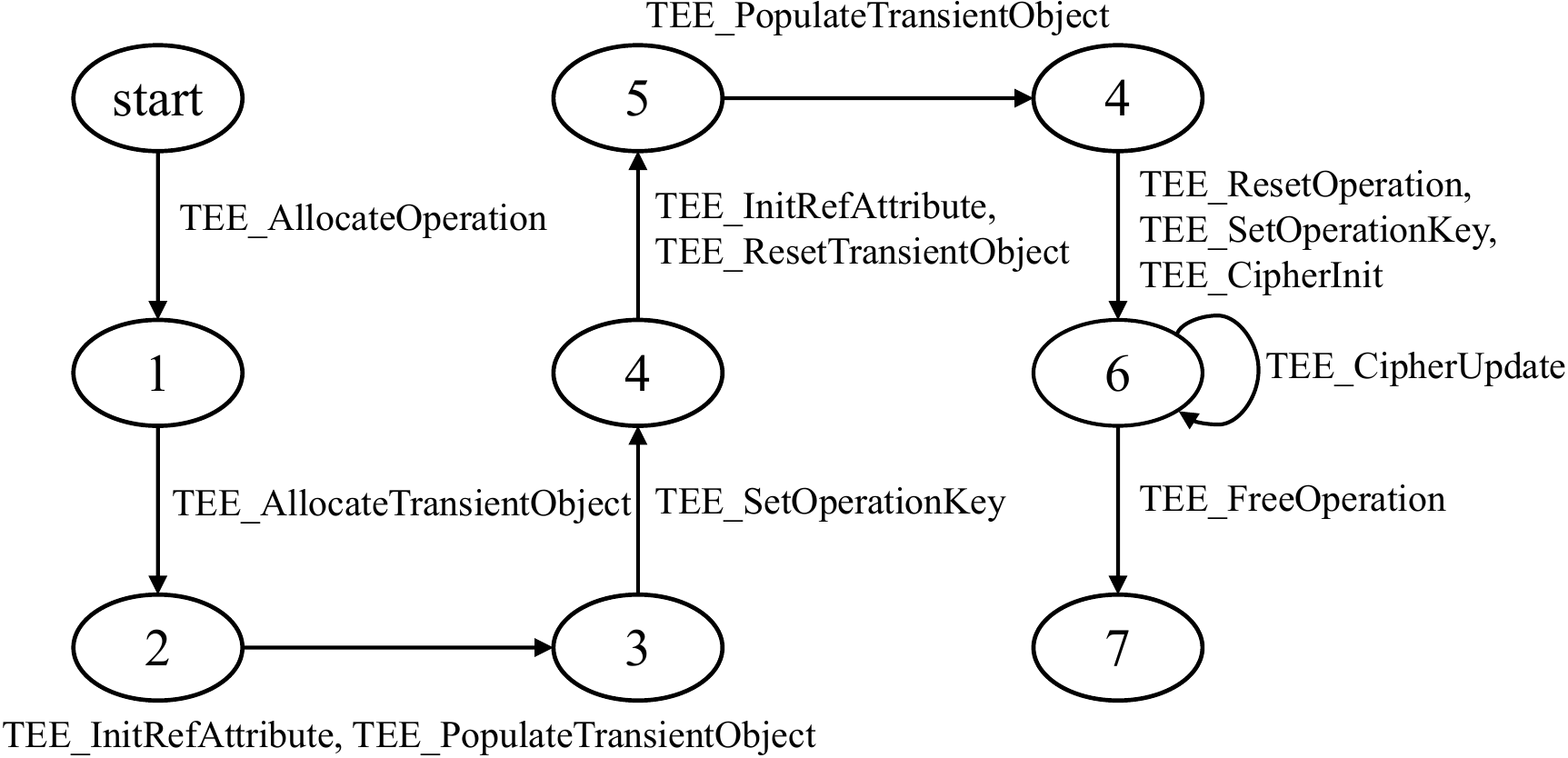} 
  \caption{An example of state transition path from the constructed state transition tree in mTower.  }
  \label{fig:stt} 
\end{figure}

To evaluate state variables' expressiveness, we construct a state transition tree to help visualize their practical meanings.
As shown in Figure~\ref{fig:stt}, we present an example state transition path from our constructed state transition tree.
We replace the state hashes with numerical values in the node label to enhance the readability.
Below is the meaning of the example state transition tree.
First, \texttt{TEE\_OperationHandle} and \texttt{TEE\_ObjectHandle} are allocated by \texttt{TEE\_AllocateOperation} and \texttt{TEE\_AllocateTriansientObject}, respectively.
Once a handle is allocated, the state transition is triggered.
Second, \texttt{TEE\_ObjectHandle} loads a dummy key by executing \texttt{TEE\_InitRefAttribute} and \texttt{TEE\_PopulateTransientObject}.
The dummy key is then loaded into the \texttt{TEE\_OperationHandle} by \texttt{TEE\_SetOperationKey}.
Then state node 4 is triggered, indicating mTower is in the \texttt{key\_set} state, which is consistent with the GP TEE Internal Core API specification.
Third, \texttt{TEE\_ObjectHandle} is reset and \texttt{TEE\_OperationHandle} loads an encryption key by executing the same syscalls of the second procedure with a valid key.
Fourth, mTower turns to a new state after \texttt{TEE\_CipherInit}, which is specified as \texttt{key\_set \& initialized} state in the specification.
This state means that mTower loads the key and the initialization vector for encryption.
When mTower is in the \texttt{key\_set \& initialized}, mTower processes the ciphering operation on provided buffers by \texttt{TEE\_CipherUpdate}.
Finally, mTower turns to a new state after it frees all the encryption configurations by \texttt{TEE\_FreeOperation}.
As a result, our state presentation mechanism truthfully reflects the workflow of symmetric encryption.

\vspace{-1mm}
\begin{mybox}
  \textbf{RQ3:}  
  On average, our active state variable inference method is effective and achieves 83.3\% precision on average.
  In addition, from the state transition tree, the inferred state variables are meaningful. 
\end{mybox}

\vspace{-1mm}
\subsection{Real World Trusted OSes (RQ4)}\label{sec:overallperformance}
\vspace{-2mm}
We apply \system on mTower from Samsung, TinyTEE from Tsinglink Cloud, and Link TEE Air for 90 hours.
The results are shown in Table~\ref{tab:longtime-results}.
The branch and state coverage explored by \system on Link TEE Air is relatively low because Link TEE Air has a more complicated and large code base.
As for vulnerability detection, a total of \bugCount{} vulnerabilities are found by \system, and \bugconfirm{} of them are confirmed, while vendors are still investigating the remaining bug reports.
We have reported confirmed vulnerabilities to MITRE, and \cve{} of them have already been assigned CVEs.

\begin{table}[tbp]
  \centering
  \caption{The number of unique vulnerabilities, branches and states found by \system in 90 hours. }
    \resizebox{0.80\linewidth}{!}{%
  \begin{tabular}{cccc}
  \toprule
  Target & Unique bugs & Branches & States \\
  \midrule
  mTower  & \mtowerbug{} & \mtowercov{} & \mtowerstate{} \\
  TinyTEE & \tinyteebug{} & \tinyteecov{} & \tinyteestate{} \\
  Link TEE Air & \linkteebug{} & \linkteecov{} & \linkteestate{} \\
  \bottomrule
  \end{tabular}
  }
  \vspace{-2mm}
  \label{tab:longtime-results}
\end{table}

We categorize vulnerabilities into seven types following \textit{the Common Weakness Enumeration (CWE) List} \cite{cwe} and we have the following conclusions. 
First, the Trusted OSes suffer from frequent null/untrusted pointer dereference vulnerabilities.
They lack code for validating supplied input pointers. When a TA tries to read or write a malformed pointer by invoking Trusted OS syscalls, a crash will be triggered, resulting in a DoS attack.
Even worse, carefully designed pointers provided by a TA to the syscall can compromise the integrity of the execution context and lead to arbitrary code execution, posing a significant risk.
The problem is severe since Trusted OSes frequently rely on pointers to transfer complex data structures, which are used in cryptographic operations.
Second, the Trusted OSes allocate resources without checking bounds. 
They allow a TA to achieve excessive memory allocation via a large $len$ value.
Since the IoT TEEs are resource-limited, these vulnerabilities may easily cause denial of service. 
Third, the Trusted OSes suffer from buffer overflow vulnerabilities.
They missed checks for buffer accesses with incorrect length values and allowed a TA to trigger a memory overwriting, DoS, and information disclosure. 
We discuss the mitigation in Section~\ref{sec:discuss} and list the vulnerabilities with their root cause analysis found on mTower and TinyTEE in Table~\ref{tab:bugslist} in Appendix~\ref{app:litedbugs}.

\textbf{Case study 1: buffer overflow (CVE-2022-35858).} 
\system{} identifies a stack-based buffer overflow vulnerability in \texttt{TEE\_PopulateTransientObject} syscall in mTower, which has 7.8 CVSS Score according to \textit{CVE Details}\cite{cvedetails}.
Specifically, the \texttt{TEE\_PopulateTransientObject} syscall creates a local array directly using the parameter \texttt{attrCount} without checking its size.
Once \texttt{TEE\_PopulateTransientObject} is invoked with a large number in \texttt{attrCount}, a memory overwrite will be triggered, resulting in Denial-of-Service (DoS), information leakage, and arbitrary code execution.

\textbf{Case study 2: null pointer dereference (CVE-2022-40759).}
\system{} uncovers a null pointer dereference in \texttt{TEE\_MACCompareFinal} in mTower.
A DoS attack can be triggered by invoking \texttt{TEE\_MACCompareFinal} with a null pointer for the parameter operation.
This bug is hard to trigger by traditional fuzzing, as it requires testing on a specific syscall sequence to enter a specific state.
\system{} can effectively identify such syscall sequences that trigger new states and prioritize testing them.
In this way, this syscall sequence can be fuzzed to cause a null pointer dereference vulnerability; otherwise, it will be discarded.

\begin{mybox}
  \textbf{RQ4:}  
  mTower, TinyTEE and Link TEE Air are all vulnerable.
  \system identifies \mtowerbug{} vulnerabilities on mTower, \tinyteebug{} vulnerabilities on TinyTEE, and \linkteebug{} vulnerabilities on Link TEE Air, resulting in \cve{} CVEs. 
\end{mybox}

%%%%%%%%%%%%%%%%%%%%%%%%%%%%%%%%%%%%%%%%%%%%%%%%%%%%%%%%%%%%%%%%%%%%%%%%%%%%%%%%
\section{Discussion}
\label{sec:discuss}
%%%%%%%%%%%%%%%%%%%%%%%%%%%%%%%%%%%%%%%%%%%%%%%%%%%%%%%%%%%%%%%%%%%%%%%%%%%%%%%%
\vspace{-2mm}
\textbf{Ethic.} We pay special attention to the potential ethical issues in this work.
First, we obtain all the tested Trusted OSes from legitimate sources.
Second, we have made responsible disclosure to report all vulnerabilities to IoT vendors.

\textbf{Lessons.} 
Based on our evaluations, we provide several security insights about the existing popular IoT Trusted OSes.
First, mTower and TinyTEE are similar to OP-TEE, an open-source TEE implementation designed for Cortex-A series MCUs, and they have several similar vulnerabilities.
For instance, they both implement vulnerable syscalls \texttt{TEE\_Malloc} and \texttt{TEE\_Realloc}, which allow an excessive memory allocation via a large value. 
In a resource-constrained MCU, such implementations can cause a crash and result in a DoS attack.
Thus, the security principles should be rethought to meet the requirements of the IoT scenario.
As a mitigation, we suggest adding checks when allocating memory; this suggestion is adopted by Samsung and TsingLink Cloud.
Second, for the null/untrusted pointer dereference and buffer overflow vulnerabilities, the input pointers and critical parameters should be especially carefully checked.
For the untrusted pointer dereferences on handlers, in addition to checking if the pointer address is valid, we suggested adding a handler list to mark if they are allocated or released.
Third, we found mTower and TinyTEE implement several critical syscalls provided by Trusted OS in non-privileged codes, which exposes a number of null/untrusted pointer dereference vulnerabilities.
Thus, a TA can easily reach these syscalls and damage the Trusted OSes.
Even worse, mTower and TinyTEE do not have memory protection mechanisms, such as ASLR (Address Space Layout Randomization).
Trusted OSes and TAs are all loaded into the same fixed address in the virtual address space.
The above two problems make the exploitation of Trusted OSes easier.
However, implementing privileged codes and memory protection mechanisms requires additional overhead, which may be unacceptable for IoT devices.
We suggest that the downstream TA developers should be aware of it and carefully design their TAs to mitigate this security risk, or lightweight Trusted OS implementations should be designed to minimize the overhead brought by security mechanisms. 

\textbf{Limitations and future work.}
While \system{} provides an effective way to fuzz IoT Trusted OSes, it also exposes some opportunities for future research.
First, our current prototype of \system{} primarily targets Trusted OSes following GP TEE Internal Core API specification. It has built-in support for alternative Trusted OSes, including proprietary ones, requiring certain modifications and configurations. We demonstrated this flexibility by extending to a proprietary Trusted OS and plan to extend SyzTrust for broader applicability.
Second, \system{} assumes that a TA can be installed in the Trusted OS for assisting fuzzing and ARM ETM is enabled for collecting the traces. In the case of certain Trusted OSes, such as ISEE-M, which are developed and used within a relatively tight supply chain, we will need to engage with the providers of these Trusted OSes to help assess the security of their respective Trusted OSes.
Finally, \system{} targets the Trusted OS of TEE, leaving several security aspects of TEEs to be studied, e.g., TAs and the interaction mechanism between peripherals and the Trusted OSes.

%%%%%%%%%%%%%%%%%%%%%%%%%%%%%%%%%%%%%%%%%%%%%%%%%%%%%%%%%%%%%%%%%%%%%%%%%%%%%%%%
\section{Related Work}
%%%%%%%%%%%%%%%%%%%%%%%%%%%%%%%%%%%%%%%%%%%%%%%%%%%%%%%%%%%%%%%%%%%%%%%%%%%%%%%%
\textbf{TEE vulnerability detection.}
Several researchers have studied and exploited vulnerabilities in TrustZone-based TEEs.
Marcel \textit{et al.} reverse-engineer HUAWEI's TEE components and present a critical security review \cite{busch2020unearthing}.
Some researchers studied the design vulnerabilities of TEE components\cite{ryan2019hardware, zhao2023uvscan}, such as Samsung's TrustZone keymaster \cite{shakevsky2022trust} and the interaction between the secure world and the normal world \cite{machiry2017boomerang}.
Recently, Cerdeira \textit{et al.} analyzed more than 200 bugs in TrustZone-assisted TEEs and presented a systematic view \cite{cerdeira2020sok}.
Since those works require manual efforts for vulnerability detection or only provide an automatic tool to target a specific vulnerability, some literature works on automatically testing the TEEs.
Some works try to apply other analysis tools, e.g., utilizing concolic execution \cite{busch2020finding} and fuzzing.
A number of TA fuzzing tools are developed, such as TEEzz\cite{buschteezz}, PartEmu\cite{harrison2020partemu}, Andrey's work \cite{andreylaunching} and Slava's work \cite{QualcommTruztZoneAppsFuzzing}.
On the contrary, Trusted OS fuzzing receives little attention.
To the best of our knowledge, the only tool OP-TEE Fuzzer \cite{OPTEEFuzzer} is for open-source TEEs, which is not applicable to closed-source IoT TEEs.

  \textbf{ETM-based fuzzing.} Firmware analysis primarily adopts two approaches: on-device testing \cite{liu2021ifizz, liu2023iot} and rehosting \cite{peng2020usbfuzz, clements2020halucinator}. For on-device testing, a few ETM-assisted analysis methods have been proposed for ARM platforms \cite{armdebugging, ning2021revisiting}.
  For instance, Ninjia \cite{ning2017ninja} utilizes ETM to analyze malware transparently. HApper~\cite{xue2021happer} and NScope \cite{zhou2022ncscope} utilize ETM to unpack Android applications and analyze the Android native code. 
  Recently, two studies have integrated ETM features into fuzzing projects. One is AFL++ CoreSight mode \cite{amored}, which targets the applications running on ARM Cortex-A platforms. 
  Another is $\mu$AFL \cite{li2022mu} to fuzz Linux peripheral drivers.
  However, these works have different purposes.
AFL++ CoreSight mode and $\mu$AFL focus on the application, whereas \system{} focuses on Trusted OSes for IoT devices and incurs many challenges due to the constrained resource and inability to instrumentation.
  Moreover, \system{} proposes a novel fuzzing framework and the state-aware feature to effectively test the stateful Trusted OSes.
  Consequently, \system{} is a novel hardware-assisted fuzzing approach proposed in this paper.

\textbf{State-aware fuzzing.}
Recently, state-aware fuzzing has emerged and gained the attention of the research community.
To understand the internal states of target systems, 
existing studies utilize the response code of protocol servers \cite{pham2020aflnet, gascon2015pulsar} or apply model learning \cite{comparetti2009prospex, wang2021mpinspector} to identify the server's states.
StateInspector \cite{mcmahon2022closer} utilizes explicit protocol packet sequences and run-time memory to infer the server's state machine.
However, they are not applicable to software and OSes since software and OSes do not have such response codes or packet sequences.
IJON \cite{aschermann2020ijon} proposes an annotation mechanism that allows the user to infer the states during the fuzzing procedure manually.
After that, numbers of studies work on automatically inferring the state of software and OSes \cite{wang2020typestate, fioraldi2021use, natella2022stateafl}, such as StateFuzz \cite{281444}, SGFUZZ \cite{ba2022stateful}, and FUZZUSB \cite{kim2022fuzzusb}.
However, these studies either require the source codes of targets or precise dynamic instrumentation tools.
Even worse, their targets are Linux kernels, protocols, and drivers, and their intuitions and observations are not suitable for IoT Trusted OSes.

%%%%%%%%%%%%%%%%%%%%%%%%%%%%%%%%%%%%%%%%%%%%%%%%%%%%%%%%%%%%%%%%%%%%%%%%%%%%%%%%
\section{Conclusion}
%%%%%%%%%%%%%%%%%%%%%%%%%%%%%%%%%%%%%%%%%%%%%%%%%%%%%%%%%%%%%%%%%%%%%%%%%%%%%%%%
We present \system{}, the first automated and practical fuzzing system to fuzz Trusted OSes for IoT devices.
We evaluate the effectiveness of the \system{} on the Nuvoton M2351 board with a Cortex M23, and the results show \system{} outperforms the baseline fuzzer \syzkaller{} with \higherCodeCov{}\% higher code coverage, \higherStateCov{}\% higher state coverage, and \higherBug{}\% improved vulnerability-finding capability.
Furthermore, we apply \system{} to evaluate real-world IoT Trusted OSes from three leading IoT vendors and detect \bugCount{} previously unknown vulnerabilities with security impacts.
In addition, we present the understanding of Trusted OS vulnerabilities and discuss the limitation and future work.
We believe \system{} provides developers with a powerful tool to thwart TEE-related vulnerabilities within modern IoT devices and complete the current TEE fuzzing scope.

\section*{Acknowledgments}
We sincerely appreciate our shepherd and all the anonymous reviewers for their insightful comments. 
This work was partly supported by NSFC under No. U1936215, the State Key Laboratory of Computer Architecture (ICT, CAS) under Grant No. CARCHA202001, the Fundamental Research Funds for the Central Universities (Zhejiang University NGICS Platform), SNSF PCEGP2\_186974, ERC StG 850868, Google Scholar Award, Meta Faculty Award, and China Scholarship Council.

\bibliographystyle{IEEEtranS}
\bibliography{Ref}

%%%%%%%%%%%%%%%%%%%%%%%%%%%%%%%%%%%%%%%%%%%%%%%%%%%%%%%%%%%%%%%%%%%%%%%%%%%%%%%%
\newpage % The Meta-Review should at least start on a new column
\appendices
%%%%%%%%%%%%%%%%%%%%%%%%%%%%%%%%%%%%%%%%%%%%%%%%%%%%%%%%%%%%%%%%%%%%%%%%%%%%%%%%

\section{The Motivation of the New Scheduling}\label{app:motivation}

\begin{figure}[tbp]
  \centering
   \includegraphics[width=0.5\textwidth]{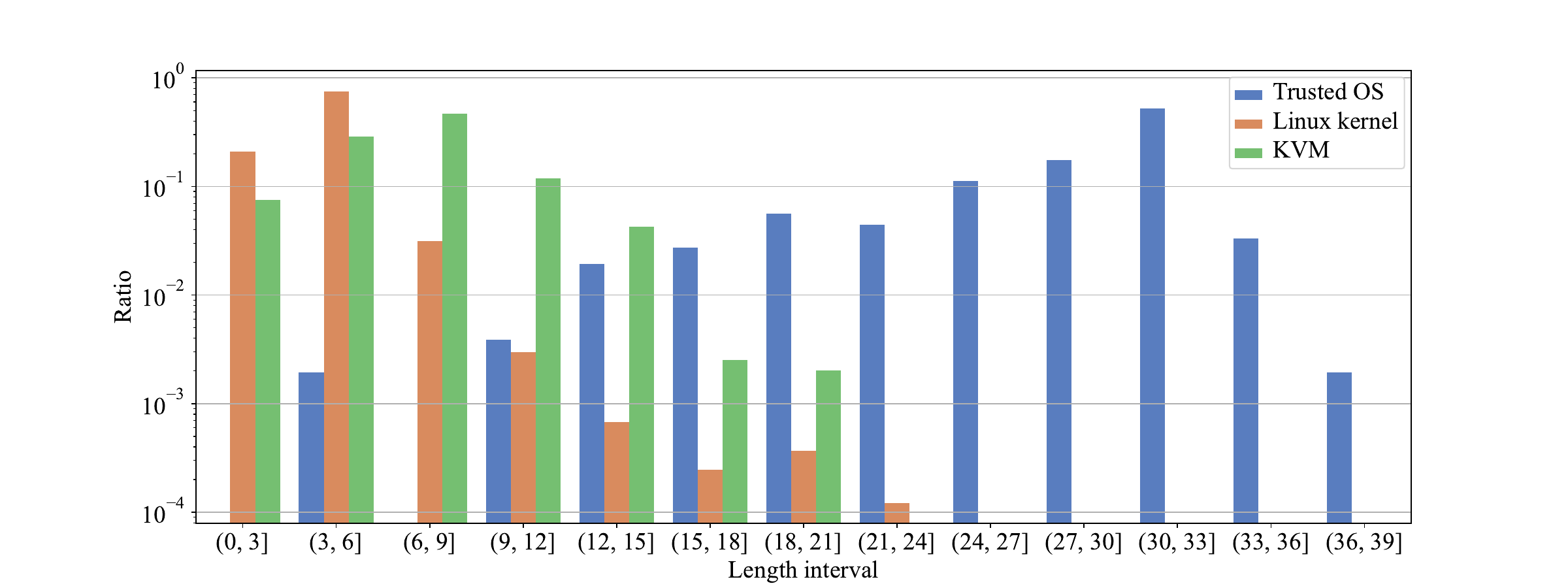} 
  \caption{The seed length ratio.}
  \label{fig:seedlen} 
\end{figure}

To justify our motivation for the new scheduling tasks design for fuzzing the IoT Trusted OS implementation, we compare the seed length ratio when testing Trusted OSes, Linux kernel and KVM.
In the evaluation, we utilize \syzkaller{} to test mTower, Linux kernel (git checkout 356d82172), and KVM v5.19 for 24 hours, and the results are shown in Figure~\ref{fig:seedlen}.
As shown in Figure~\ref{fig:seedlen}, the average seed length when fuzzing Trusted OSes is 27.8 while the others are less than 7.
In the triage scheduling task, \syzkaller{} removes syscalls one by one in a syscall sequence.
Then \syzkaller{} tests the modified syscall sequences to get the smallest syscall sequence that maintains the same code coverage.
For those syscall sequences whose length is more than 30, the triage scheduling tasks probably spend lots of time on removing syscalls and testing the modified syscall sequences.
In addition, we count the number of minimized syscalls when fuzzing mTower.
In a 48-hours fuzzing, only 56 syscall sequences are minimized, and among them, 18 syscall sequences only are removed with one syscall in the triaging tasks.
Thus, \system{} doesn't have to perform triaging tasks since most test cases will not be minimized.

\section{Scope and Scalability of \system{}}\label{app:assumption}
We first provide an overview of the major Trusted OSes from leading IoT vendors and use the objective data to validate our assumptions.
Then, we justify how to extend \system{} for Cortex-A TEE OSes.

\begin{table}[htbp]
  \centering
  \caption{An overview of the major Trusted OS implementations provided by leading IoT vendors. }
  \resizebox{\linewidth}{!}{%
    \begin{tabular}{cccccc}
      \toprule
      Vendor          & Trusted OS        & Standards          & \begin{tabular}[c]{@{}c@{}}Support \\         (installing TA)\end{tabular} & Some of supported devices                &  \\ \midrule
      Samsung         & mTower            & GP Standards       & \fullcirc                    & NuMaker-PFM-M2351                         &  \\
      Alibaba         & Link TEE Air      & Proprietary                 & \fullcirc                    & NuMaker-PFM-M2351 &  \\
      TsingLink Cloud & TinyTEE           & GP Standards       & \fullcirc                    & NuMaker-PFM-M2351/LPC55S69/STM32L562      &  \\
      Beanpod         & ISEE-M            & GP Standards       & \fullcirc                    & LPC55S series/GD32W515/STM32L5 series     &  \\
      Trustonic       & Kinibi-M          & PSA Certified APIs & \fullcirc                    & MicroChip SAML11                          &  \\
      ARM             & TF-M & PSA Certified APIs & \fullcirc                    & NuMaker-PFM-M2351, STM32L5, ...     & \\ \bottomrule
      \end{tabular}
  }
  \label{tab:iottees}
\end{table}

\begin{table}[htbp]
  \centering
  \caption{ETM feature on IoT devices. }
  \resizebox{\linewidth}{!}{%
  \begin{tabular}{cccc}
    \toprule
    Manufacturer       & Device & \begin{tabular}[c]{@{}c@{}}Privilige Secure Debug \\         (including ETM)\end{tabular} & \begin{tabular}[c]{@{}c@{}}Debug Authentication\\ Managerment\end{tabular} \\ \midrule
    Nuvoton            & NuMaker-PFM-M2351                    & Enable in default                                                                                                 & ICP programming tool                                                                               \\
    NXP Semiconductors & LPC55S69                             & Enable in default                                                                                                 & Debug credential certificate                                                                       \\
    STMicroelectronics & STM32L562                            & Enable in default                                                                                                 & STM32CubeProgrammer                                                                                \\
    GigaDevice         & GD32W515                             & Enable in default                                                                                                 & Efuse                                                                                              \\
    MicroChip          & SAML11                               & Enable in default                                                                                                 & Extern debugger            \\ \bottomrule                                                                       
    \end{tabular}
  }
  \label{tab:iotetm}
\end{table}

As shown in Table~\ref{tab:iottees}, there are six major Trusted OSes for IoT devices, which are widely adopted by the major IoT MCUs \cite{mTower, TinyTEE, tinyteeapp, LinkTEE,Kinibim,tfm,isee-m}.
Three of them follow the GP standards, and they all allow TA installation.
In addition, as shown in Table~\ref{tab:iotetm}, though the devices have multiple debug authentication to disable the privilege secure debug (debug in secure privileged modes), they enable privilege secure debug by default, thereby enabling the ETM feature to collect execution traces.
Thus, \system{} can be directly deployed on half of the major Trusted OSes.
As for other Trusted OSes, \system{} can be applied with modification introduced in Section~\ref{sec:impl}.

Though our paper focuses on Cortex-M Trusted OS for IoT devices,
\system{} can be applied to those Trusted OSes for Cortex-A that allows installing a TA and having their ETM feature enabled.
For instance, OP-TEE from Linaro, Link TEE Pro from Ali Cloud, and iTrustee from Huawei meet these assumptions.
However, for the Cortex-A Trusted OSes that are developed and employed in a relatively private supply chain, such as QSEE from Qualcomm, we can collaborate with those mobile vendors and help them vet the security of their Trusted OSes.

\section{Vulnerabilites Found by \system{} }\label{app:litedbugs}
We present all the vulnerabilities found by \system{} on mTower, TinyTEE and Link TEE Air with their root cause in Table~\ref{tab:bugslist}. We update the vulnerabilities disclosure progress in the GitHub link \url{https://github.com/SyzTrust}.

\begin{table*}[tbp]
\renewcommand\arraystretch{1.2}
  \centering
\normalsize
\caption{Vulnerabilities detected by \system{}.}\label{tab:bugslist}
\resizebox{\linewidth}{!}{%
\begin{tabular}{@{}lllll@{}}
  \toprule
  Vul. ID & Target       & Description                                                                                     & Status                                                              & Root Cause \& Impact                                                                                                                                                                                                                                                                                         \\ \midrule
  1       & mTower       & \begin{tabular}[c]{@{}l@{}}Allocation of resources \\ without limits or throttling\end{tabular} & \begin{tabular}[c]{@{}l@{}}CVE-2022-38155\\ (7.5 HIGH)\end{tabular} & TEE\_Malloc allows a TA to achieve excessive memory allocation via a large len value                                                                                                                                                                                                                         \\
  2       & mTower       & \begin{tabular}[c]{@{}l@{}}Allocation of resources \\ without limits or throttling\end{tabular} & \begin{tabular}[c]{@{}l@{}}CVE-2022-40762\\ (7.5 HIGH)\end{tabular} & TEE\_Realloc allows a TA to achieve excessive memory allocation via a large len value                                                                                                                                                                                                                        \\
  3       & mTower       & \begin{tabular}[c]{@{}l@{}}Allocation of resources \\ without limits or throttling\end{tabular} & Confirmed                                                           & TEE\_AllocateOperation allows a TA to achieve excessive memory allocation via a large len value                                                                                                                                                                                                              \\
  4       & mTower       & \begin{tabular}[c]{@{}l@{}}Allocation of resources \\ without limits or throttling\end{tabular} & Confirmed                                                           & TEE\_AllocateTransientObject allows a TA to achieve excessive memory allocation via a large len value                                                                                                                                                                                                        \\
  5       & mTower       & Improper input validation                                                                       & \begin{tabular}[c]{@{}l@{}}CVE-2022-40761\\ (7.5 HIGH)\end{tabular} & \begin{tabular}[c]{@{}l@{}}The function tee\_obj\_free allows a TA to trigger a denial of service by invoking the function TEE\_AllocateOperation with a disturbed heap layout, related \\ to utee\_cryp\_obj\_alloc\end{tabular}                                                                            \\
  6       & mTower       & Buffer overflow                                                                                 & Reported                                                            & \begin{tabular}[c]{@{}l@{}}A Buffer Access with Incorrect Length Value vulnerability in the TEE\_MemMove function allows a TA to trigger a  denial of service by invoking the function \\ TEE\_MemMove with a "size" parameter that exceeds the size of "dest".\end{tabular}                                 \\
  7       & mTower       & Buffer overflow                                                                                 & \begin{tabular}[c]{@{}l@{}}CVE-2022-40760\\ (7.5 HIGH)\end{tabular} & \begin{tabular}[c]{@{}l@{}}A Buffer Access with Incorrect Length Value vulnerability in the TEE\_MACUpdate function allows a TA to trigger a denial of service by invoking the function \\ TEE\_MACUpdate with an excessive size value of chunkSize.\end{tabular}                                            \\
  8       & mTower       & Buffer overflow                                                                                 & \begin{tabular}[c]{@{}l@{}}CVE-2022-40757\\ (7.5 HIGH)\end{tabular} & \begin{tabular}[c]{@{}l@{}}A Buffer Access with Incorrect Length Value vulnerability in the TEE\_MACComputeFinal function allows a TA to trigger a denial of service by invoking \\ the function TEE\_MACComputeFinal with an excessive size value of messageLen.\end{tabular}                               \\
  9       & mTower       & Buffer overflow                                                                                 & \begin{tabular}[c]{@{}l@{}}CVE-2022-40758\\ (7.5 HIGH)\end{tabular} & \begin{tabular}[c]{@{}l@{}}A Buffer Access with Incorrect Length Value vulnerability in the TEE\_CipherUpdate function allows a TA to trigger a denial of service by invoking the function \\ TEE\_CipherUpdate with an excessive size value of srcLen.\end{tabular}                                         \\
  10      & mTower       & Buffer overflow                                                                                 & Reported                                                            & \begin{tabular}[c]{@{}l@{}}A Buffer Access with Incorrect Length Value vulnerability in the TEE\_DigestDoFinal function allows a TA to trigger a denial of service by invoking the function \\ TEE\_DigestDoFinal with an excessive size value of chunkLen.\end{tabular}                                     \\
  11      & mTower       & Buffer overflow                                                                                 & Reported                                                            & \begin{tabular}[c]{@{}l@{}}A Buffer Access with Incorrect Length Value vulnerability in the TEE\_DigestUpdate function allows a TA to trigger a denial of service by invoking the function \\ TEE\_DigestUpdate with an excessive size value of chunkLen.\end{tabular}                                       \\
  12      & mTower       & \begin{tabular}[c]{@{}l@{}}Missing release of memory\\  after effective lifetime\end{tabular}   & \begin{tabular}[c]{@{}l@{}}CVE-2022-35858\\ (7.8 HIGH)\end{tabular} & \begin{tabular}[c]{@{}l@{}}The TEE\_PopulateTransientObject and \_\_utee\_from\_attr functions allow a TA to trigger a memory overwrite, denial of service, and information disclosure \\ by invoking the function TEE\_PopulateTransientObject with a large number in the parameter attrCount.\end{tabular} \\
  13      & mTower       & NULL pointer dereference                                                                        & \begin{tabular}[c]{@{}l@{}}CVE-2022-40759\\ (7.5 HIGH)\end{tabular} & TEE\_MACCompareFinal contains a NULL pointer dereference on the parameter operation                                                                                                                                                                                                                          \\
  14      & mTower       & NULL pointer dereference                                                                        & \begin{tabular}[c]{@{}l@{}}CVE-2022-36621\\ (7.5 HIGH)\end{tabular} & TEE\_AllocateTransientObject contains a NULL pointer dereference on the parameter object                                                                                                                                                                                                                     \\
  15      & mTower       & NULL pointer dereference                                                                        & \begin{tabular}[c]{@{}l@{}}CVE-2022-36622\\ (7.5 HIGH)\end{tabular} & TEE\_GetObjectInfo1 contains a NULL pointer dereference on the parameter objectInfo                                                                                                                                                                                                                          \\
  16      & mTower       & NULL pointer dereference                                                                        & Confirmed                                                           & TEE\_GetObjectInfo contains a NULL pointer dereference on the parameter objectInfo                                                                                                                                                                                                                           \\
  17      & mTower       & Untrusted pointer dereference                                                                   & Reported                                                            & Uncertain (provided the PoC to the vendor)                                                                                                                                                                                                                                                                   \\
  18      & mTower       & Untrusted pointer dereference                                                                   & Reported                                                            & Uncertain (provided the PoC to the vendor)                                                                                                                                                                                                                                                                   \\
  19      & mTower       & Untrusted pointer dereference                                                                   & Reported                                                            & \begin{tabular}[c]{@{}l@{}}TEE\_GetObjectInfo and utee\_cryp\_obj\_get\_info functions allow a corruption on the link field of object handle and then a Denial of Service (DoS) will \\ be triggered by invoking the function tee\_obj\_get\end{tabular}                                                     \\
  20      & mTower       & Untrusted pointer dereference                                                                   & Reported                                                            & An invalid pointer dereference can be triggered when a TA tries to read a malformed TEE\_OperationHandle by the TEE\_MACComputeFinal function                                                                                                                                                                \\
  21      & mTower       & Untrusted pointer dereference                                                                   & Reported                                                            & An invalid pointer dereference can be triggered when a TA tries to read a malformed TEE\_OperationHandle by the TEE\_SetOperationKey2 function                                                                                                                                                               \\
  22      & mTower       & Untrusted pointer dereference                                                                   & Reported                                                            & An invalid pointer dereference can be triggered when a TA tries to read a malformed TEE\_OperationHandle by the TEE\_MACUpdate function                                                                                                                                                                      \\
  23      & mTower       & Untrusted pointer dereference                                                                   & Reported                                                            & An invalid pointer dereference can be triggered when a TA tries to read a malformed TEE\_OperationHandle by the TEE\_GetOperationInfoMultiple function                                                                                                                                                       \\
  24      & mTower       & Untrusted pointer dereference                                                                   & Reported                                                            & An invalid pointer dereference can be triggered when a TA tries to read a malformed TEE\_OperationHandle by  the TEE\_AEEncryptFinal function                                                                                                                                                                \\
  25      & mTower       & Untrusted pointer dereference                                                                   & Reported                                                            & An invalid pointer dereference can be triggered when a TA tries to read a malformed TEE\_OperationHandle by the TEE\_MACInit function                                                                                                                                                                        \\
  26      & mTower       & Untrusted pointer dereference                                                                   & Reported                                                            & An invalid pointer dereference can be triggered when a TA tries to read a malformed TEE\_OperationHandle by the TEE\_SetOperationKey function                                                                                                                                                                \\
  27      & mTower       & Untrusted pointer dereference                                                                   & Reported                                                            & An invalid pointer dereference can be triggered when a TA tries to read a malformed TEE\_OperationHandle by the TEE\_ResetOperation function                                                                                                                                                                 \\
  28      & mTower       & Untrusted pointer dereference                                                                   & Reported                                                            & An invalid pointer dereference can be triggered when a TA tries to read a malformed TEE\_OperationHandle by the TEE\_DigestUpdate function                                                                                                                                                                   \\
  29      & mTower       & Untrusted pointer dereference                                                                   & Reported                                                            & An invalid pointer dereference can be triggered when a TA tries to read a malformed TEE\_OperationHandle by the TEE\_AEDecryptFinal function                                                                                                                                                                 \\
  30      & mTower       & Untrusted pointer dereference                                                                   & Reported                                                            & An invalid pointer dereference can be triggered when a TA tries to read a malformed TEE\_OperationHandle by the TEE\_CipherInit function                                                                                                                                                                     \\
  31      & mTower       & Untrusted pointer dereference                                                                   & Reported                                                            & An invalid pointer dereference can be triggered when a TA tries to read a malformed TEE\_OperationHandle by the TEE\_FreeOperation function                                                                                                                                                                  \\
  32      & mTower       & Untrusted pointer dereference                                                                   & Reported                                                            & An invalid pointer dereference can be triggered when a TA tries to read a malformed TEE\_OperationHandle by the TEE\_DigestDoFinal function                                                                                                                                                                  \\
  33      & mTower       & Untrusted pointer dereference                                                                   & Reported                                                            & An invalid pointer dereference can be triggered when a TA tries to read a malformed TEE\_OperationHandle by the TEE\_AllocateOperation function                                                                                                                                                              \\
  34      & mTower       & Untrusted pointer dereference                                                                   & Reported                                                            & An invalid pointer dereference can be triggered when a TA tries to read a malformed TEE\_OperationHandle by the TEE\_FreeTransientObject function                                                                                                                                                            \\
  35      & mTower       & Untrusted pointer dereference                                                                   & Reported                                                            & An invalid pointer dereference can be triggered when a TA tries to read a malformed TEE\_OperationHandle by the TEE\_AEUpdate function                                                                                                                                                                       \\
  36      & mTower       & Untrusted pointer dereference                                                                   & Reported                                                            & An invalid pointer dereference can be triggered when a TA tries to read a malformed TEE\_OperationHandle by the TEE\_ResetOperation function                                                                                                                                                                 \\
  37      & mTower       & Untrusted pointer dereference                                                                   & Reported                                                            & Uncertain (provided the PoC to the vendor)                                                                                                                                                                                                                                                                   \\
  38      & mTower       & Untrusted pointer dereference                                                                   & Reported                                                            & Uncertain (provided the PoC to the vendor)                                                                                                                                                                                                                                                                   \\
  39      & TinyTEE      & \begin{tabular}[c]{@{}l@{}}Allocation of resources \\ without limits or throttling\end{tabular} & Confirmed                                                           & TEE\_Malloc allows a trusted application to achieve Excessive Memory Allocation via a large len value                                                                                                                                                                                                        \\
  40      & TinyTEE      & \begin{tabular}[c]{@{}l@{}}Allocation of resources \\ without limits or throttling\end{tabular} & Confirmed                                                           & TEE\_Realloc allows a trusted application to achieve Excessive Memory Allocation via a large len value                                                                                                                                                                                                       \\
  41      & TinyTEE      & NULL pointer dereference                                                                        & Confirmed                                                           & TEE\_AllocateTransientObject contains a NULL pointer dereference on the parameter object                                                                                                                                                                                                                     \\
  42      & TinyTEE      & Untrusted pointer dereference                                                                   & Confirmed                                                           & An invalid pointer dereference can be triggered when a TA tries to read a malformed TEE\_OperationHandle by the TEE\_DigestUpdate function                                                                                                                                                                   \\
  43      & TinyTEE      & Untrusted pointer dereference                                                                   & Confirmed                                                           & An invalid pointer dereference can be triggered when a TA tries to read a malformed TEE\_OperationHandle by the TEE\_SetOperationKey function                                                                                                                                                                \\
  44      & TinyTEE      & Untrusted pointer dereference                                                                   & Confirmed                                                           & An invalid pointer dereference can be triggered when a TA tries to read a malformed TEE\_OperationHandle by the TEE\_SetOperationKey function                                                                                                                                                                \\
  45      & TinyTEE      & Untrusted pointer dereference                                                                   & Confirmed                                                           & An invalid pointer dereference can be triggered when a TA tries to read a malformed TEE\_OperationHandle by the TEE\_ResetOperation function                                                                                                                                                                 \\
  46      & TinyTEE      & Untrusted pointer dereference                                                                   & Confirmed                                                           & An invalid pointer dereference can be triggered when a TA tries to read a malformed TEE\_OperationHandle by the TEE\_FreeOperation function                                                                                                                                                                  \\
  47      & TinyTEE      & Untrusted pointer dereference                                                                   & Confirmed                                                           & An invalid pointer dereference can be triggered when a TA tries to read a malformed TEE\_OperationHandle by the TEE\_CipherDoFinal function                                                                                                                                                                  \\
  48      & TinyTEE      & Untrusted pointer dereference                                                                   & Confirmed                                                           & An invalid pointer dereference can be triggered when a TA tries to read a malformed TEE\_OperationHandle by the TEE\_CipherInit function                                                                                                                                                                     \\
  49      & TinyTEE      & Untrusted pointer dereference                                                                   & Confirmed                                                           & An invalid pointer dereference can be triggered when a TA tries to read a malformed TEE\_OperationHandle by the TEE\_AllocateOperation function                                                                                                                                                              \\
  50      & TinyTEE      & Untrusted pointer dereference                                                                   & Confirmed                                                           & An invalid pointer dereference can be triggered when a TA tries to read a malformed TEE\_OperationHandle by the TEE\_AsymmetricSignDigest function                                                                                                                                                           \\
  51      & TinyTEE      & Untrusted pointer dereference                                                                   & Confirmed                                                           & An invalid pointer dereference can be triggered when a TA tries to read a malformed TEE\_OperationHandle by the TEE\_CipherUpdate function                                                                                                                                                                   \\
  52      & Link TEE Air & NULL pointer dereference                                                                        & Reported                                                            & tee\_memcpy calls tee\_osa\_memcpy, which contains a NULL pointer dereference on the result object                                                                                                                                                                                                           \\
  53      & Link TEE Air & NULL pointer dereference                                                                        & Reported                                                            & tee\_strcpy calls tee\_osa\_strcpy, which contains a NULL pointer dereference on the result object                                                                                                                                                                                                           \\
  54      & Link TEE Air & NULL pointer dereference                                                                        & Reported                                                            & tee\_memset calls tee\_osa\_strcpy, which contains a NULL pointer dereference on the result object                                                                                                                                                                                                           \\
  55      & Link TEE Air & Buffer overflow                                                                                 & Reported                                                            & \begin{tabular}[c]{@{}l@{}}tee\_hash\_update does not check the size of its second parameter "size" and calls dev\_ioctl, which triggers an invalid memory access with large "size" value \\ when consecutively copying 64 bytes in tee\_osa\_memcpy\end{tabular}                                            \\
  56      & Link TEE Air & NULL pointer dereference                                                                        & Reported                                                            & tee\_memmove calls tee\_osa\_memmove, which contains a NULL pointer dereference on the result object                                                                                                                                                                                                         \\
  57      & Link TEE Air & NULL pointer dereference                                                                        & Reported                                                            & tee\_strcat calls tee\_osa\_strcat, which contains a NULL pointer dereference on the result object                                                                                                                                                                                                           \\
  58      & Link TEE Air & Buffer overflow                                                                                 & Reported                                                            & \begin{tabular}[c]{@{}l@{}}tee\_hash\_digest does not check the size of its third parameter "size" and calls dev\_ioctl, which triggers an heap overflow with large "size" value and causes \\ an invalid pointer dereference in tee\_osa\_free\end{tabular}                                                 \\
  59      & Link TEE Air & Untrusted pointer dereference                                                                   & Reported                                                            & \begin{tabular}[c]{@{}l@{}}An invalid pointer dereference can be triggered when a trusted application tries to access an invalid address in the pool\_free function called \\ by tee\_osa\_free and tee\_free\end{tabular}                                                                                   \\
  60      & Link TEE Air & NULL pointer dereference                                                                        & Reported                                                            & tee\_strncpy calls tee\_osa\_strncpy, which contains a NULL pointer dereference on the result object                                                                                                                                                                                                         \\
  61      & Link TEE Air & NULL pointer dereference                                                                        & Reported                                                            & tee\_strcasecmp calls tee\_osa\_strcasecmp, which contains a NULL pointer dereference on the s1 and s2 object                                                                                                                                                                                                \\
  62      & Link TEE Air & NULL pointer dereference                                                                        & Reported                                                            & tee\_memcmp calls tee\_osa\_memcmp, which contains a NULL pointer dereference on the s1 and s2 object                                                                                                                                                                                                        \\
  63      & Link TEE Air & Untrusted pointer dereference                                                                   & Reported                                                            & An invalid pointer dereference can be triggered when a trusted application tries to access an invalid address in the tee\_hash\_init function                                                                                                                                                                \\
  64      & Link TEE Air & NULL pointer dereference                                                                        & Reported                                                            & tee\_strncat calls tee\_osa\_strncat, which contains a NULL pointer dereference on the result object                                                                                                                                                                                                         \\
  65      & Link TEE Air & NULL pointer dereference                                                                        & Reported                                                            & tee\_strnlen calls tee\_osa\_strnlen, which contains a NULL pointer dereference on the s object                                                                                                                                                                                                              \\
  66      & Link TEE Air & Buffer overflow                                                                                 & Reported                                                            & \begin{tabular}[c]{@{}l@{}}tee\_base64\_encode does not check the size of its parameters "src\_len" and "dst", which triggers a buffer overflow if "src\_len" is larger than the size \\ of "dst" and ruins the metadata of the next chunk\end{tabular}                                                      \\
  67      & Link TEE Air & NULL pointer dereference                                                                        & Reported                                                            & tee\_strlen calls tee\_osa\_strlen, which contains a NULL pointer dereference on the s object                                                                                                                                                                                                                \\
  68      & Link TEE Air & NULL pointer dereference                                                                        & Reported                                                            & tee\_strcmp calls tee\_osa\_strcmp, which contains a NULL pointer dereference on the s1 and s2 object                                                                                                                                                                                                        \\
  69      & Link TEE Air & Buffer overflow                                                                                 & Reported                                                            & \begin{tabular}[c]{@{}l@{}}tee\_hash\_final does not check the size of its first parameter "dgst" and calls dev\_ioctl, which triggers an heap overflow if the size of "dgst" is smaller than \\ 48 bytes and causes an invalid pointer dereference in tee\_osa\_free\end{tabular}                           \\
  70      & Link TEE Air & Untrusted pointer dereference                                                                   & Reported                                                            & \begin{tabular}[c]{@{}l@{}}An invalid pointer dereference can be triggered when a trusted application tries to access an invalid address in the tee\_osa\_memcpy function \\ called by tee\_aes\_init\end{tabular}                                                                                           \\ \bottomrule
\end{tabular}
  }
  \end{table*}

  %%%%%%%%%%%%%%%%%%%%%%%%%%%%%%%%%%%%%%%%%%%%%%%%%%%%%%%%%%%%%%%%%%%%%%%%%%%%%%%%
  \section{Meta-Review}
  %%%%%%%%%%%%%%%%%%%%%%%%%%%%%%%%%%%%%%%%%%%%%%%%%%%%%%%%%%%%%%%%%%%%%%%%%%%%%%%%
  \subsection{Summary}
  This paper presents a fuzzing framework for TEEs on IoT devices. It leverages hardware-assisted features such as Arm ETM to collect traces. It uses the state and code coverage as composite feedback to guide the fuzzer to effectively explore more states.

\subsection{Scientific Contributions}
\begin{itemize}
\item Creates a New Tool to Enable Future Science
\item Identifies an Impactful Vulnerability
\item Provides a Valuable Step Forward in an Established Field
\end{itemize}

\subsection{Reasons for Acceptance}
\begin{enumerate}
\item Creates a New Tool to Enable Future Science. This paper presents a new fuzzing tool that targets IoT Trusted OSes.
\item Identifies an Impactful Vulnerability. Several vulnerabilities were identified, and CVEs are disclosed.
\item Provides a Valuable Step Forward in an Established Field. This paper leverages hardware-assisted features such as Arm ETM to further improve the effectiveness of TEE OS fuzzing on IoT devices.
\end{enumerate}

\subsection{Noteworthy Concerns} % Exclude if your meta-review does not have noteworthy concerns
The proposed fuzzing framework targets Trusted OSes following GP TEE internal API specification. It is assumed that a TA can be installed in the trusted OS for assisting fuzzing. Arm ETM needs to be enabled for collecting the traces.

%%%%%%%%%%%%%%%%%%%%%%%%%%%%%%%%%%%%%%%%%%%%%%%%%%%%%%%%%%%%%%%%%%%%%%%%%%%%%%%%
\section{Response to the Meta-Review} % Optional
%%%%%%%%%%%%%%%%%%%%%%%%%%%%%%%%%%%%%%%%%%%%%%%%%%%%%%%%%%%%%%%%%%%%%%%%%%%%%%%%
The meta-review notes that our proposed fuzzing framework, SyzTrust, targets trusted OSes following GP TEE Internal API specification. To clarify, SyzTrust also has built-in support for testing alternative Trusted OSes, including proprietary ones, as demonstrated by our extension of SyzTrust to the proprietary "Link TEE Air". We are also extending our SyzTrust prototype to other OSes.

The meta-review notes that SyzTrust assumed that a TA can be installed in the trusted OS for assisting fuzzing and ARM ETM needs to be enabled for collecting the traces. We agree and note that these assumptions align with typical IoT Trusted OS scenarios. SyzTrust focuses on the Trusted OS binaries provided by IoT vendors, delivering security insights for device manufacturers and end users. First, given that IoT device manufacturers often need to implement device-specific TAs, Trusted OS binaries supplied by IoT vendors generally allow TA installation (similar to smartphones where manufacturers can similarly install TAs in the respective TEEs). Second, SyzTrust tests IoT Trusted OSes by deploying them on development boards where ETM is enabled by default. For certain Trusted OSes that are developed and used within a relatively private supply chain, we will need to engage with the providers of these Trusted OSes to help assess the security of their respective Trusted OSes.
Appendix~\ref{app:assumption} provides a detailed discussion along with supporting data.

\end{document}